\documentclass[prb,twocolumn,showpacs,superscriptaddress,preprintnumbers,amsmath,amssymb]{revtex4-1}
 \usepackage{graphicx,color}
 \usepackage{dcolumn}% Align table columns on decimal point
 \usepackage{bm}% bold math
 \usepackage{epstopdf}
 \usepackage{tabularx}

\begin{document}

%\normalem	% changes \emph back to normal after introducing ulem package.

\title{Ultrafast spin dynamics: role of laser-induced modification of exchange parameters}

\author{Sergiy Mankovsky}
\affiliation{Department of Chemistry/Phys. Chemistry, LMU Munich,
Butenandtstrasse 11, D-81377 Munich, Germany}

\author{Svitlana Polesya}
\affiliation{Department of Chemistry/Phys. Chemistry, LMU Munich,
  Butenandtstrasse 11, D-81377 Munich, Germany}

\author{Hubert Ebert}
\affiliation{Department of Chemistry/Phys. Chemistry, LMU Munich,
  Butenandtstrasse 11, D-81377 Munich, Germany}


\date{\today}

\begin{abstract}
 Induced by an ultra-short laser pulse, the electronic
  structure of a material undergoes strong modifications  leading to a fast
demagnetization in magnetic materials. Induced spin-flip 
transitions are one of the reasons for demagnetization,
that is discussed in the literature as a Stoner-like mechanism.
On the other hand, demagnetization due to transverse spin fluctuations
is usually discussed on the basis of the Heisenberg Hamiltonian and
hardly accounts for the modification of the electronic structure.
In this work we demonstrate a strong impact of the laser-induced
electron transitions, both spin-flip and spin-conserving, on the
exchange coupling parameters. For this, a simple two-step scheme is
suggested. As a first step, the electronic structure time
evolution during the ultra-short laser pulse is described accurately
within time-dependent density-functional theory (TD-DFT) calculations. As a next step, the 
information on the time-dependent electronic structure is used for
calculations of the parameters of the Heisenberg Hamiltonian. A strong
modification of the exchange coupling parameters is found in response
to the applied ultra-short laser pulse. The most important reason for this
modification is played by the laser induced repopulation of the 
electronic states.
Although the changes of the exchange parameters are most prominent
during the laser pulse, they may be important
also for the magnetic relaxation. The same concerns the spin-lattice
interactions playing a central role for the relaxation process.
A strong impact of the laser-induced modification of the electronic
structure on the spin-lattice coupling parameters is also shown in this work.
\end{abstract}

\maketitle

\section{INTRODUCTION}

Ultrafast spin manipulation is one of the most exciting research topics in
the latest decades. Since the first report by Beaurepaire et
al.\cite{BMDB96}, demonstrating that magnetic order can be manipulated 
on a sub-picosecond timescale using ultrashort laser pulses, many efforts
have been done, both experimental and theoretical, to shed light on the
physical processes behind ultrafast magnetic dynamics on the one hand
side and to adjust these processes to control, e.g. the magnetization
dynamics in magnetic materials 
\cite{KKR10,BV13,CBR17}. Despite the huge progress made by these
activities, the various microscopic mechanisms responsible for the
  demagnetization at different time scales is still under
  debate.\cite{BMDB96,BGBS04,FIHT15,KMD+10,HB96,FIHT15,DAS+19}
  Definitely, one can see a consensus in the literature concerning
  the role of the spin-orbit coupling (SOC) responsible
  for the laser-induced demagnetization in metals leading to an angular
  momentum transfer from the spin subsystem to other degrees 
  of freedom \cite{ZG08,FIHT15} via spin-flip transitions during the laser
  field induced excitation \cite{ZH00,ZG08} or Elliott-Yafet-like processes with spin-flip 
  electron scattering on phonons \cite{KMD+10,CBO11,IHF13,CBLO15}. 
  The Stoner-type demagnetization mechanism is described reasonably well 
  within time dependent density functional theory (TD-DFT)
  \cite{KDE+14,SSED22}.
  It should be stressed that such calculations are usually performed for a
  non-distorted crystal lattice and a collinear magnetic state.
  On the other hand, the impact of the electron-phonon scattering on the
  time evolution of the magnetization was reported recently \cite{SSED22} by
  performing TD-DFT calculations in the presence of preexcited
  phonon modes in the system.  
  Apart from TD-DFT, other quantum-mechanical investigations have been
  done estimating the rate of angular momentum
transfer from the electron to the lattice and magnetic degrees of freedom
via SOC-driven spin-flip electron scattering
\cite{KMD+10,SK13,IHF13,FIHT15,FTI+17,DAS+19}. 

The experimental observation of a strong laser-induced magnetization
decay observed  by Eich et al. \cite{EPR+17} in Co led
the authors to the conclusion, that the  ferromagnetic-paramagnetic phase
transition is a result of an extremely efficient ultrafast generation of
collective transverse spin excitations, i.e. magnons, rather than a
loss of the exchange splitting. 
Tengdin et al. \cite{TYC+18}, also report about the demagnetization in Ni 
which occurs much earlier than the collapse of the exchange splitting.
The authors, however, suggest that the ultrafast demagnetization is driven
by a highly nonequilibrium process, e.g. superdiffusive spin currents.

Attributing the ultrafast demagnetization to magnetic disorder in
a ferromagnet, a natural question is what is the efficient mechanism  
responsible for ultrafast generation of magnons, which will depend on the
material. To find the answer to this question, 
an efficient way is to combine classical models used for spin dynamics
with first-principles DFT calculations. This may be for instance the
atomistic Landau-Lifshitz-Gilbert (LLG) model based on a classical spin
Hamiltonian \cite{KNC+08} 
giving access to the magnetic torque accounting for contributions from
isotropic exchange, Dzyaloshinskii-Moriya interactions (DMI), and magnetic 
anisotropy calculated from first principles \cite{DHSN17}.  
The ab-initio parametrized quantum kinetic approach 
was reported recently to describe magnon occupation dynamics due to
electron-magnon scattering in ferromagnetic metals excited by ultrafast
laser radiation \cite{WO23}. The results demonstrate a strong demagnetization of Fe
within 200 fs due to an ultrafast generation of high-energy magnons.
The optically induced magnetic torque may be a consequence
of a coupling between the electric field of the laser light and the electron spins
\cite{NRT+12,SHR+14,MHS+15,FBM16,Men17,HGSL22}, and may originate,
e.g., due to the optical spin transfer torque \cite{NFAM04,NRT+12} or due to
the inverse Faraday effect (IFE) \cite{SHR+14,MHS+15,FBM16,Men17,HGSL22}.
Furthermore, it may be seen as a consequence of the optically-induced
changes of the interatomic exchange interactions \cite{SHR+14,MHS+15}.
The laser induced demagnetization and relaxation may also be controlled
by spin-lattice interactions, which are responsible for the angular
momentum transfer between the magnetic subsystem and the crystal lattice
leading to ultrafast magnon or phonon generation in the demagnetization
and relaxation processes, respectively \cite{HB96,RSBD20,FDC+20}.

One should note, however, that the efficiency of the magnon generation should depend not only 
on the mechanism of magnon excitation, but also on the properties of
the magnon spectrum determined by the parameters of the spin Hamiltonian. 
Ultrafast heating of the electronic subsystem should almost immediately
(i.e. on the time scale of $\sim 10^{-15}$ s) result
in a modification of all parameters entering the equations used to describe
spin-dynamics. Accordingly, all parameters should change with time
following the absorbed energy and corresponding time evolution of the
electronic subsystem. This concerns, in particular, the anisotropy
field generated due to the absorbed laser pulse \cite{KJK+02}.
In general, the theoretical description of these changes for
the non-equilibrium electron density require to go
beyond the standard approach used for the equilibrium, and should rely,
for instance, on the non-equilibrium Green’s-function formalism
\cite{SBLK13,FBM16}, or exact diagonalization approach \cite{HZ98}. 
Alternatively, we suggest in the present work a simple scheme to
demonstrate the impact of the optically excited out-of-equilibrium
electron subsystem on the demagnetization or relaxation
processes. Information about the 
laser-pulse excited electron system is obtained via TD-DFT
calculations, which are used here to determine the time evolution of the
exchange coupling and spin-lattice coupling parameters.
%In general, similar time-dependent modifications should occur also 
%for others parameters, as for instance, MCA constants.
As a result, the spin dynamic characteristics should follow the changes
of all the parameters that determine the torque on the
magnetic moments.  

Note that on a longer time scale, when the magnetization is also decreased due
to increasing magnetic disorder, its impact on the exchange coupling
parameters may be significant and should also be taken into account
\cite{DHSN17}.

\section{Results and discussion}

\subsection{Magnetic torque due to the optical heating of the electronic
subsystem}

We focus first on the impact of the transverse spin fluctuations on
the laser-induced demagnetization in ferromagnets. 
When the system is in equilibrium, these fluctuations are
reasonably well described using the extended Heisenberg spin model
Hamiltonian ${\cal H}_{\rm Heis}$ 
with the parameters calculated from first principles, i.e. the
parameters representing symmetric exchange interactions, 
Dzyaloshinskii-Moriya interactions, magnetic anisotropy, etc.
The spin evolution out of equilibrium described by the atomistic
Landau-Lifshitz-Gilbert equation \cite{BNF12}, is governed by the torque
on the magnetic moments on the sites $i$, $\vec{T}_i = \hat{s}_i \times
\vec{H}_{{\rm{eff}},i}$. The torque is determined by the external perturbation field
and effective magnetic field, $\vec{H}_{{\rm{eff}},i} = -\frac{\partial
  {\cal H}}{\partial \hat{s}_i}$, given in particular, in terms of the
spin Hamiltonian ${\cal H} = {\cal H}_{\rm Heis}$.

For a system affected by an ultrafast laser pulse, 
one can consider the magnetic torque $\Delta \vec{T}_{\rm{Heis},i}$
created due to changes of the parameters of the model 
Hamiltonian, i.e. $\Delta J_{ij}(t)$ and $\Delta \vec{D}_{ij}(t)$,
$\Delta \underline{K}_{i}(t)$ as a result of the strong laser-induced
heating of the electron subsystem.
Note that in general, the torque on the magnetic moment should account 
for the impact of the spin evolution during the whole time 
period starting with the laser pulse (see, e.g. Ref.\ [\onlinecite{BNF12}]).
However, using a simplified time-dependence for the magnetic torque,
one can derive it from the current state of the electronic
structure evolving in time.
Despite the resulting uncertainty of the numerical results, this allows 
a qualitative, semi-quantitative, analysis of the impact of the
laser-excited electronic subsystem on the transverse spin dynamics.

Considering bcc Fe with a non-distorted
crystal lattice ($T_{lat} = 0$ K) as an example, we will focus on the laser induced
changes of the isotropic exchange interactions represented by the
 parameter $\Delta J_{ij}(t)$ and the corresponding magnetic torque
$\Delta \vec{T}_i = \sum_{j} (\hat{s}_i \times \hat{s}_j)\Delta J_{ij}$, while
the DMI parameters should vanish because of the
inversion symmetry. Furthermore, the parameters of the
magnetocrystalline anisotropy (MCA) are very weak in the bulk. 

The torque determined this way, however, does not give access to the angular momentum
transfer to the spin subsystem resulting in the generation of new magnons
responsible for the transverse demagnetization.
As discussed above, magnon generation can be driven by the optically
induced magnetic torques that stem from the direct laser light interaction
with the magnetic moments \cite{MHS+15}, or from the phonon-magnon
scattering. 
%Considerin spin-lattice mechanism responsible for such a process,
%we will discuss also the properties of spin-lattice coupling parameters.
%This however, may change at longer time domain due to thermalization of
%crystal lattice.
The latter mechanism is determined by spin-lattice coupling (SLC)
DMI-like, $\vec{\cal D}^{\mu}_{ij,k}$, and MCA-like,
$\underline{K}_{i,k}^{\mu}$ parameters \cite{MPL+22,MLPE23} entering the
spin-lattice Hamiltonian ${\cal H}_{\rm{sl}}$  (see Appendix
\ref{Appendix2}), and require a finite temperature of
the lattice subsystem leading to a broken local symmetry at the atomic sites
due to a finite amplitude of the lattice fluctuations. 
As a consequence, magnetic torques created by the effective fields
${{H}^{\alpha}_{\rm{sl},i}} = - \frac{\partial   
  {{\cal H}}_{\rm{sl}}}{\partial s_i^\alpha }$ \cite{MPL+22,MLPE23,RSBD20}
may change in the case of a strong laser-induced excitation of the 
electronic subsystem, leading to a corresponding induced torque $\Delta
\vec{T}_{\rm{sl},i}(t)$ determined by the change of the SLC parameters
following the time evolution of the excited electronic system.

In order to demonstrate the impact of such a modification of the exchange
coupling parameters, we consider the system at a certain time step 
instead of performing an integration of the atomistic equations of motion.
The electronic structure for any time step found within a
TD-DFT calculations (here using the Elk code \cite{ELK}) can be used later on to
calculate the exchange coupling parameters for this time step.
We focus in this work on two types of changes in the electronic
subsystem, the spin-dependent part of the electronic potential (in
particular, the exchange field $B_{\rm xc} = \frac{1}{2}(V^\uparrow - V^\downarrow)$) and
the re-population of the electronic states, which in our 
opinion has the main impact on the exchange parameters.
The goal is to demonstrate a significant role of the laser induced
modifications of these parameters for spin dynamics, in particular for
laser-induced demagnetization dynamics.

To account for the re-population when calculating the
exchange parameters, one has to adopt the expression for the exchange
coupling parameters worked out previously \cite{LKAG87,USPW03,EM09a} for
the case of a reasonably low temperatures ($\lesssim T_C$), when 
the Fermi-Dirac occupation function $f(E)$ can be well approximated by
the step function. Taking into account the temperature or laser-pulse
induced modifications of $\underline{f}(E)$ which become significant in the high
temperature (or strong pulse fluence) regime, one obtains an expression for
the elements of the $\underline{J}_{ij}$ tensor as follows (see Appendix
\ref{Appendix1})  
\begin{widetext}
%eeeeeeeeeeeeeeeeeeeeeeeeeeeeeeeeeeeeeeeeeeeeeeeeeeeeeeeeeeeeeeeeeeeeeeeeeeeeee
\begin{eqnarray}
J_{ij}^{\alpha \beta} & = & 
 - \frac{1}{2} \bigg[\frac{1}{\pi}\, \mbox{Im}\mbox{Tr} \int^{\infty} dE\,
       \frac{d\,\underline{f}(E)}{dE} \, (E - \mu(T)) \,
       \langle \Delta V \rangle_i^\alpha \, \underline{\tau}_{ij} (E) \,
         \langle \Delta V \rangle_j^\beta \, \underline{\tau}_{ji} (E) \nonumber  \\
 &  &+  \frac{1}{\pi}\, \mbox{Im}\mbox{Tr} \int^{\infty} dE\,
                  \underline{f}(E)\, \,\langle \Delta V \rangle_i^\alpha \, \underline{\tau}_{ij} (E) \,
                    \langle \Delta V \rangle_j^\beta \, \underline{\tau}_{ji} (E) \bigg]\,,
\label{Eq:NEW_Jij}
\end{eqnarray}
%eeeeeeeeeeeeeeeeeeeeeeeeeeeeeeeeeeeeeeeeeeeeeeeeeeeeeeeeeeeeeeeeeeeeeeeeeeeeee
\end{widetext}
with $\underline{f}(E)$ the spin-dependent occupation function.
A similar expression can be derived also for the spin-lattice parameters.  
However, below we focus on the properties of the isotropic
exchange coupling parameters, $J_{ij} = \frac{1}{2}(J_{ij}^{xx} +
J_{ij}^{yy})$, and the DMI-like SLC parameters.

\subsection{Exchange coupling parameters for out of
  equilibrium situation}

As a first step we consider the system being at a
finite temperature $T$. This allows us to simulate the effect of the
optically induced re-population of the electronic states, which can be
seen as a heating of the electronic structure to a temperature in
the order of $10^4$ K (see below).
The resulting impact for the heated electronic subsystem
with temperature $T_{\rm el}$ is described by the corresponding Fermi-Dirac
distribution function.
In contrast to this, we do not account for any changes of the
electron potential, as well as thermal atomic displacements and a
deviation of the spin
moments from perfect FM ordering, i.e., we neglect at the moment their
  impact on the exchange parameters.
This simplification allows to demonstrate the impact of the heated
electronic subsystem on the magnetic properties of the system in
  the thermal equilibrium represented by the Heisenberg model. 

\begin{figure*}[t]
\includegraphics[trim={0cm 0cm 0cm
  0cm},width=0.16\textwidth,angle=0,clip]{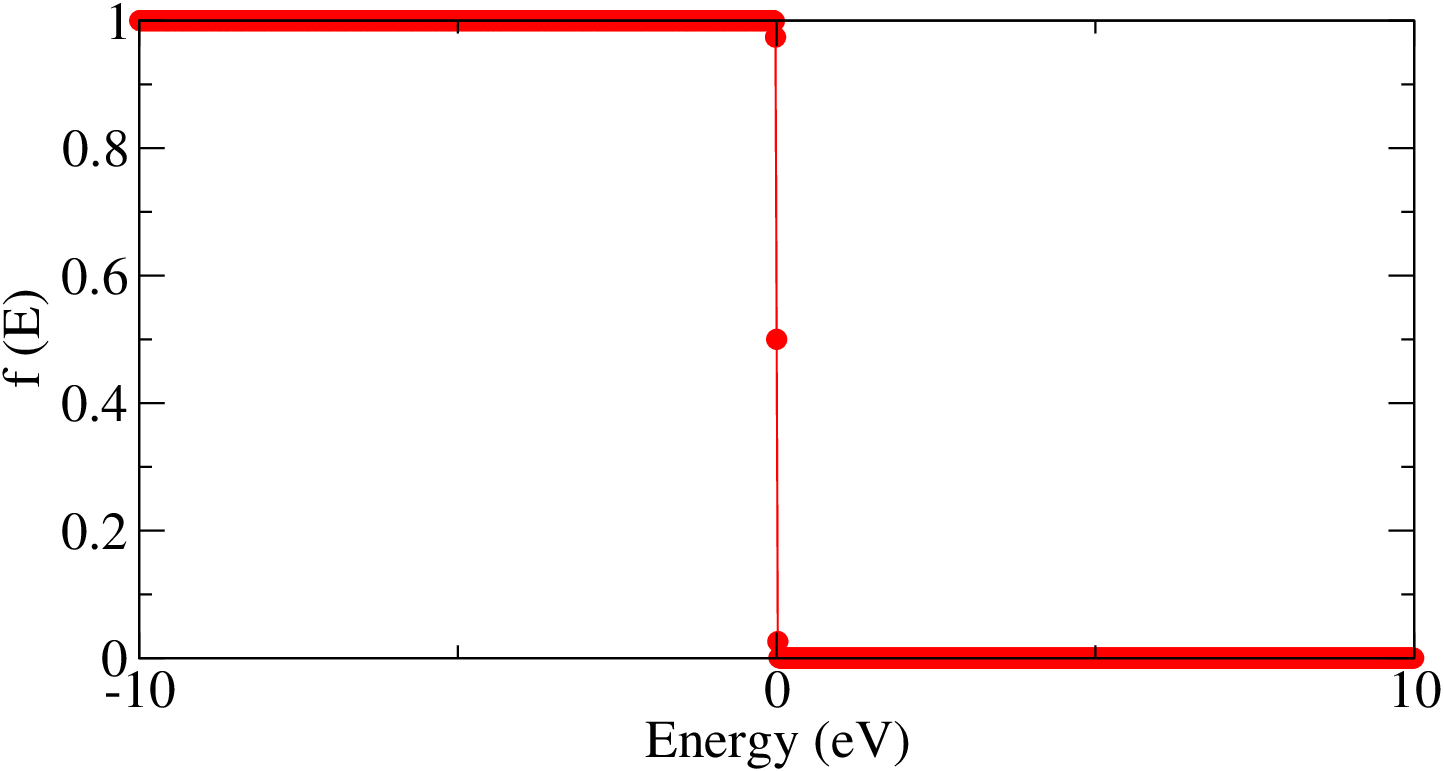}    \,\,\,\,\,\,\,\,
\includegraphics[trim={0cm 0cm 0cm
  0cm},width=0.16\textwidth,angle=0,clip]{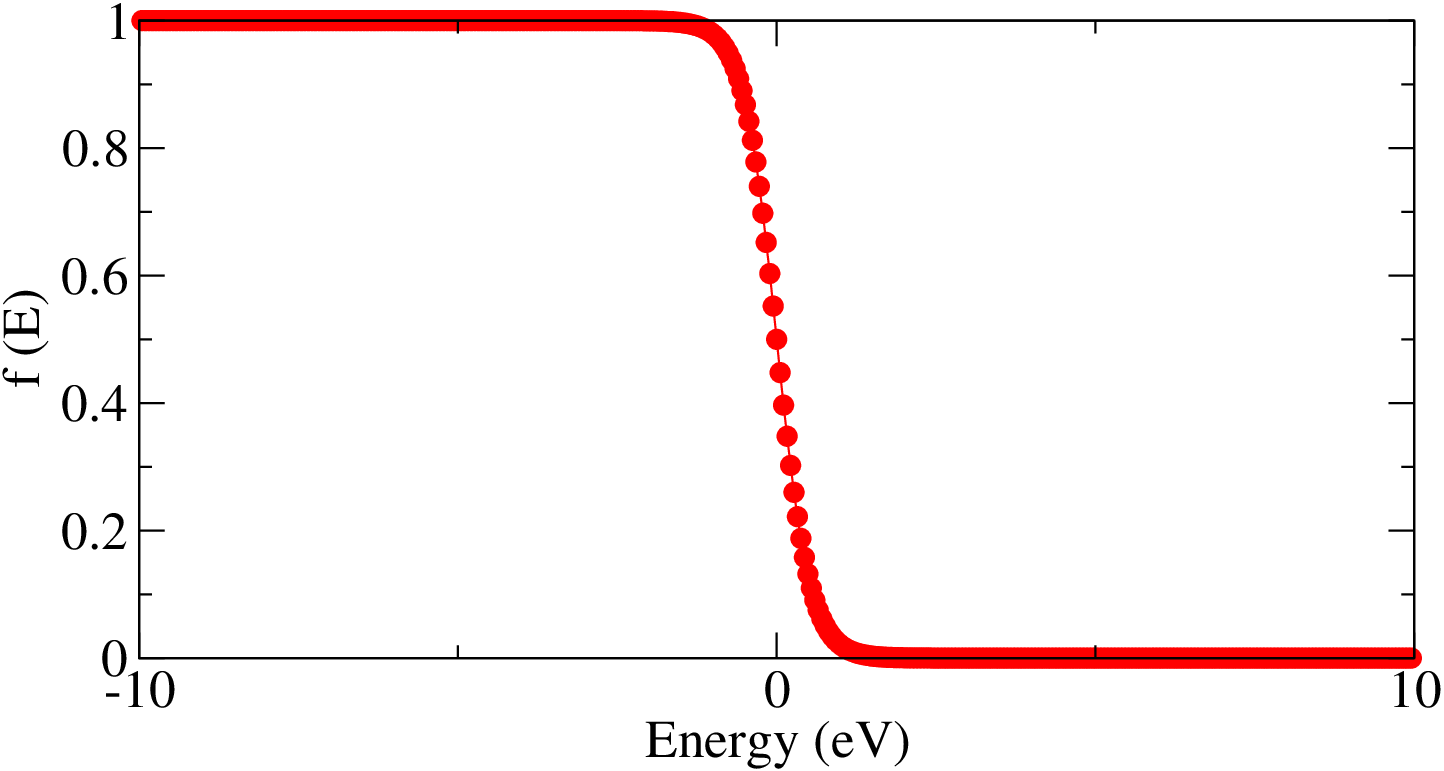} \,\,\,\,\,\,\,\,
\includegraphics[trim={0cm 0cm 0cm
  0cm},width=0.16\textwidth,angle=0,clip]{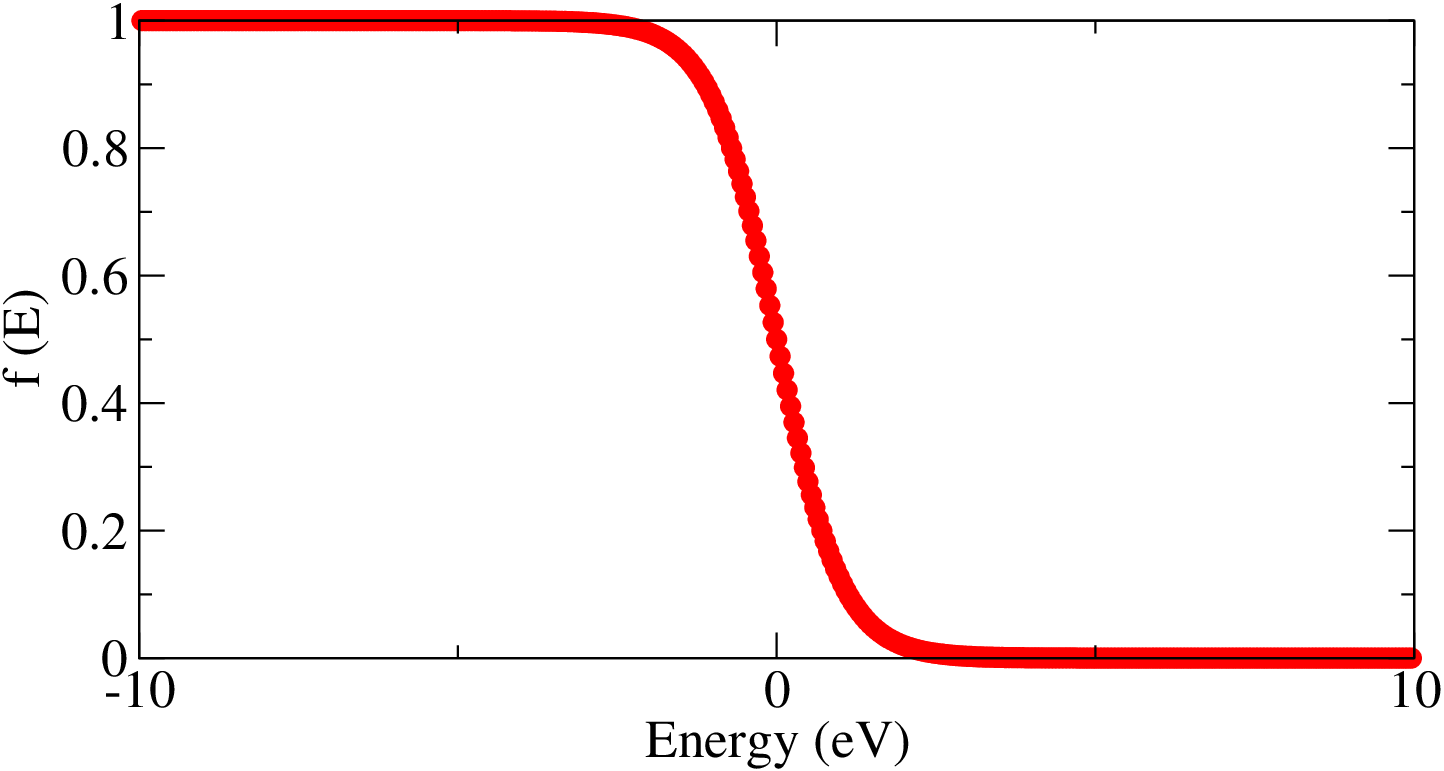} \,\,\,\,\,\,\,\,
\includegraphics[trim={0cm 0cm 0cm
  0cm},width=0.16\textwidth,angle=0,clip]{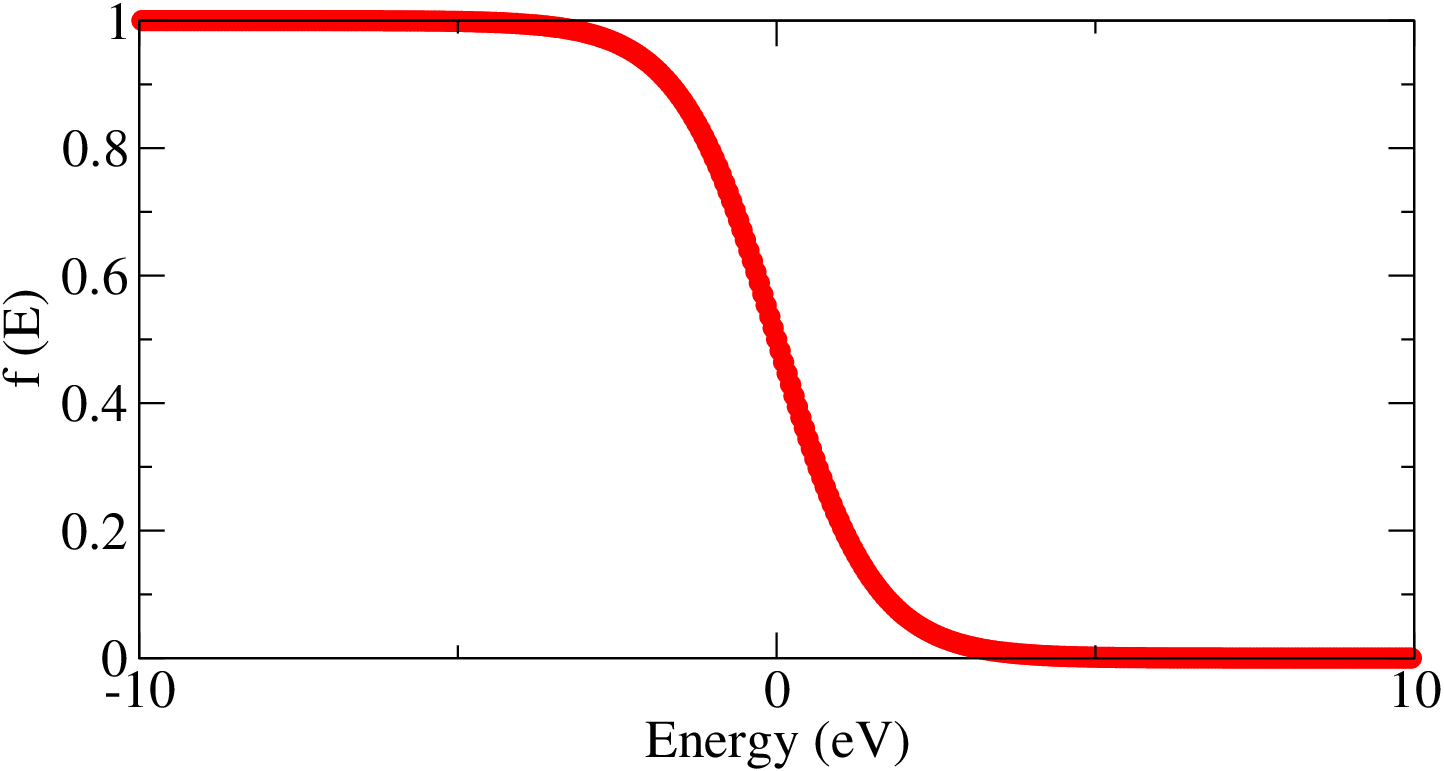} \,\,\,\,\,\,\,\,
\includegraphics[trim={0cm 0cm 0cm
  0cm},width=0.16\textwidth,angle=0,clip]{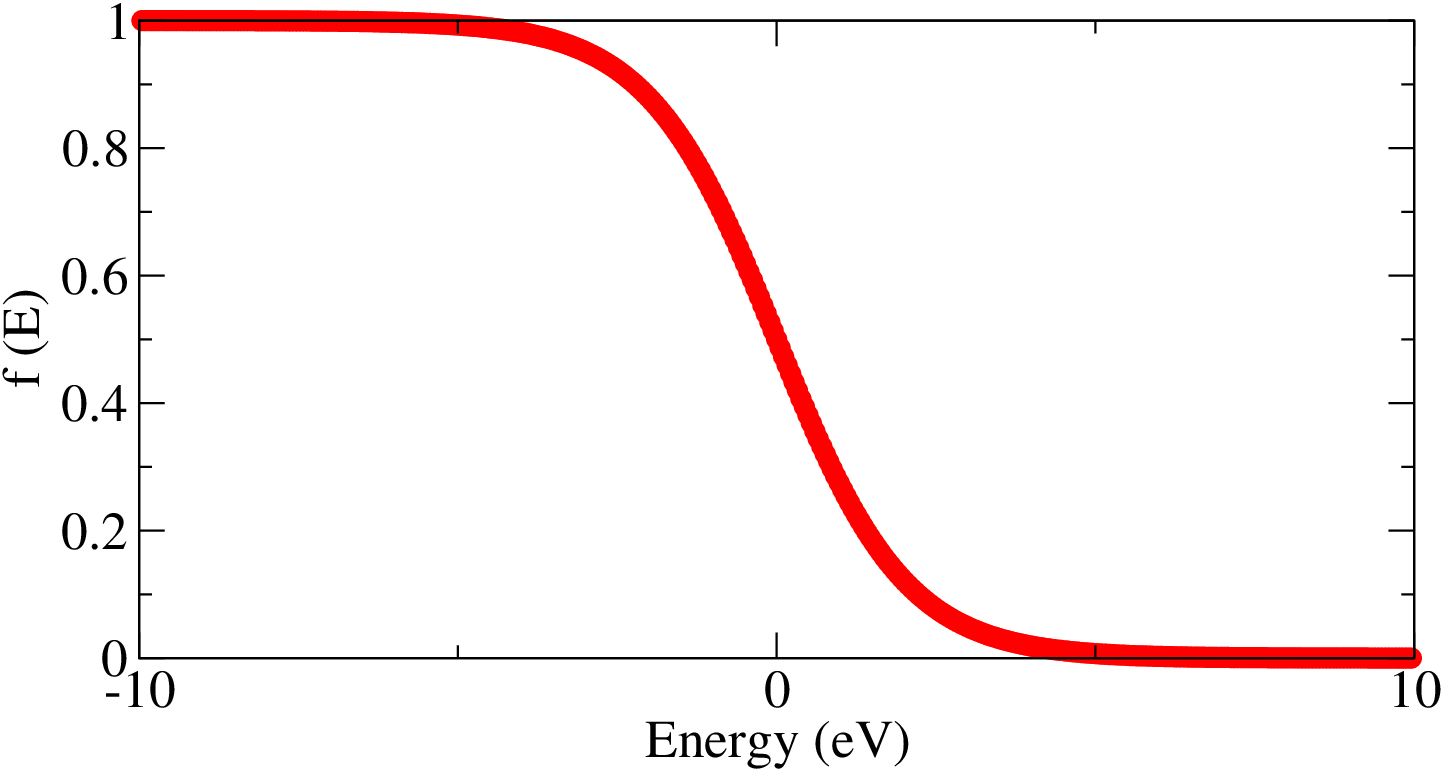} 
\includegraphics[trim={0cm 0cm 0cm
  0cm},width=0.16\textwidth,angle=0,clip]{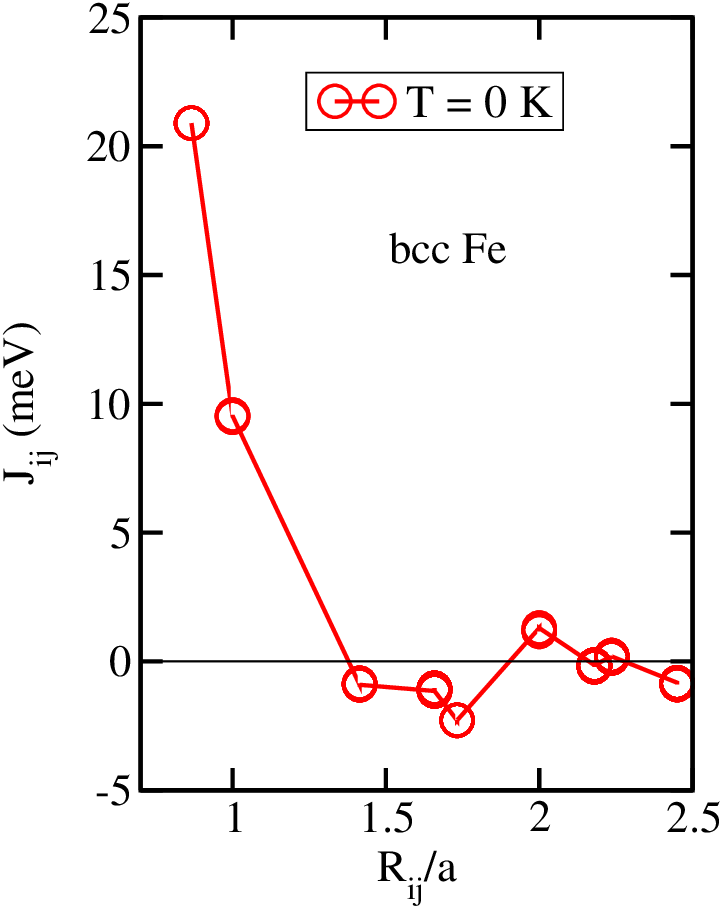}\,(a)\,
\includegraphics[trim={0cm 0cm 0cm
  0cm},width=0.16\textwidth,angle=0,clip]{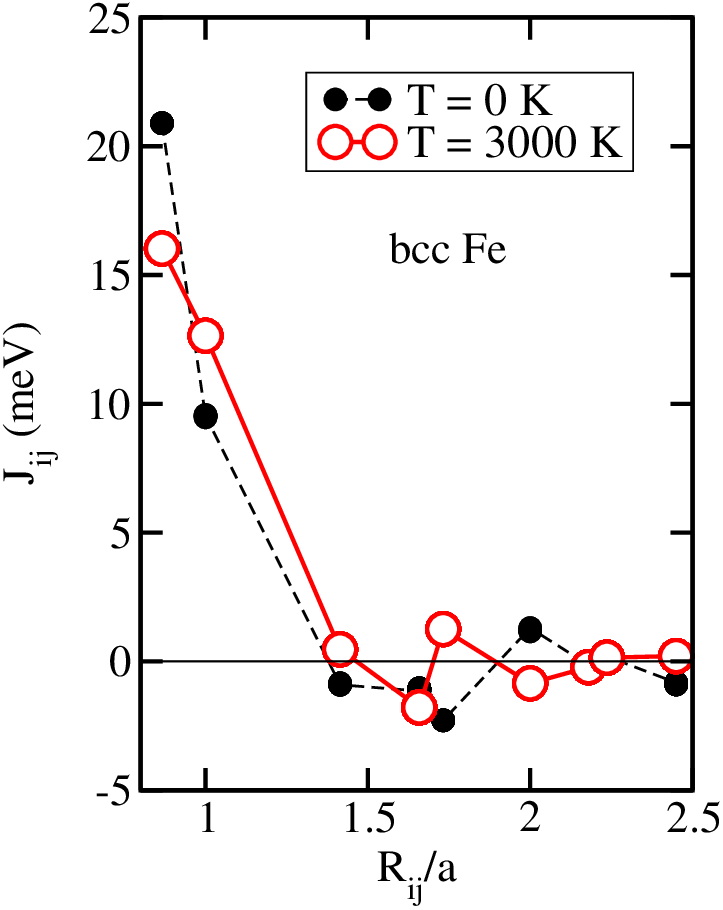}\,(b)\,
\includegraphics[trim={0cm 0cm 0cm
  0cm},width=0.16\textwidth,angle=0,clip]{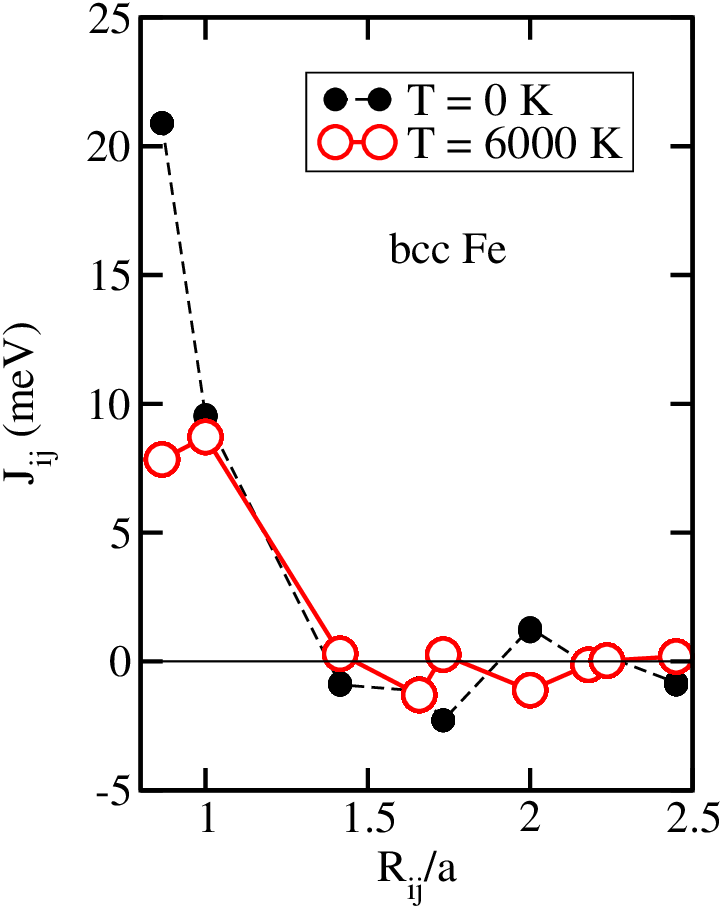}\,(c)\,
\includegraphics[trim={0cm 0cm 0cm
  0cm},width=0.16\textwidth,angle=0,clip]{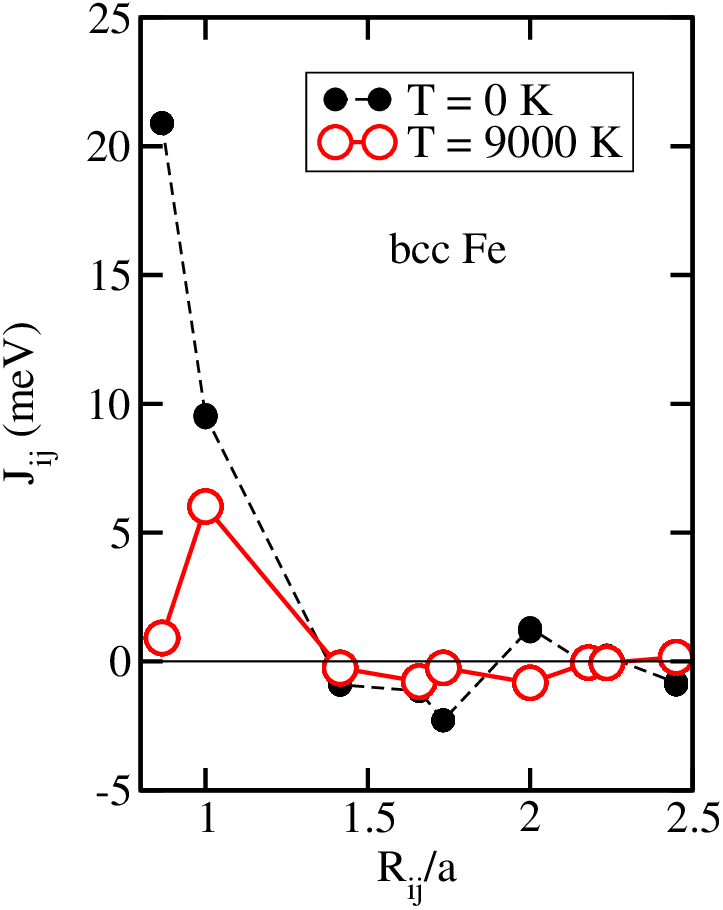}\,(d)\,
\includegraphics[trim={0cm 0cm 0cm
  0cm},width=0.16\textwidth,angle=0,clip]{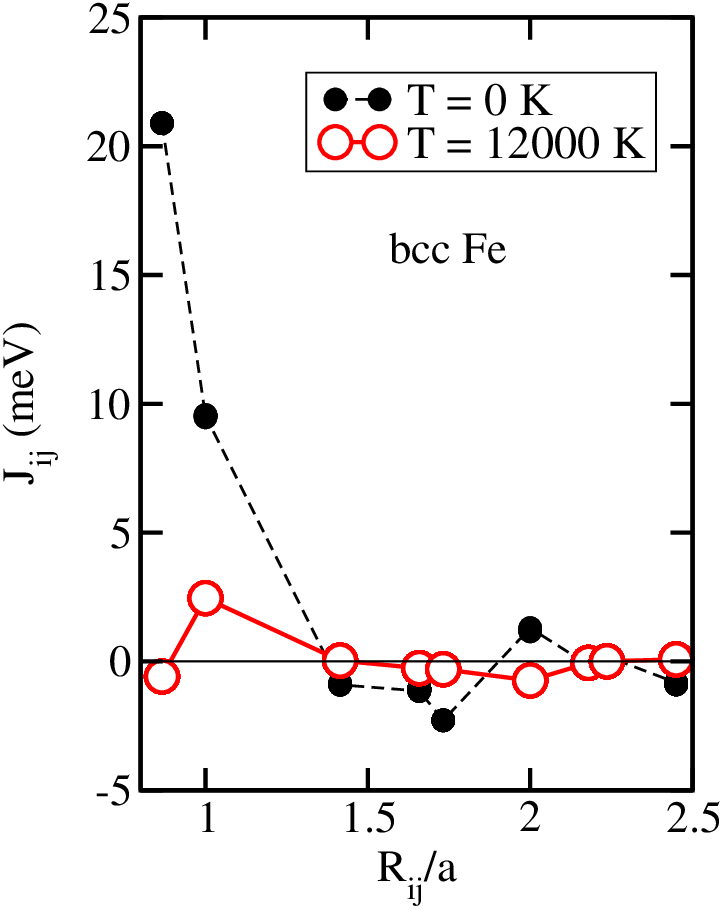}\,(e)\,
\caption{ The Fermi-Dirac distribution corresponding to
temperatures (a) 0 K, (b) 3000 K, (c) 6000 K, (d) 9000 K, (e) 12000 K,
together with corresponding results for the temperature dependent
exchange coupling parameters $J_{ij}(T)$ for bcc Fe.}
\label{fig:Fe_JXC-Temper}
\end{figure*}

The parameters $J_{ij}$ calculated for bcc Fe for five different
temperatures are represented in Fig.\ \ref{fig:Fe_JXC-Temper}. 
We focus here only on these parameters, first of all because
they are responsible for the FM order and characterize
the Curie temperature in the system, while the DMI parameters are equal
to zero for symmetry reason. 

The top panel in  Fig.\ \ref{fig:Fe_JXC-Temper} shows the temperature
dependent Fermi function $f(E,T_{\rm el})$ for the
temperatures (a) 0 K, (b) 3000 K, (c) 6000 K, (d) 9000 K, and (e) 12000 K. 
The corresponding exchange coupling parameters $J_{ij}$ for bcc Fe are shown in the bottom
panel. One can clearly see their significant decrease with 
increasing electronic temperature. This in turn should lead to a 
decrease of the Curie temperature, as can be demonstrated by
Monte Carlo simulations for a finite temperature $T_{\rm Heis}$  of the spin
subsystem. It is worth noting that this implies thermalization of the spin
subsystem, that occurs at the timescale of $\sim 10^{-12}$ s, i.e. much
longer than the duration of a typical laser pulse considered here
\cite{KKR10,CBR17}. This point has to be discussed in addition, when
dealing with a realistic situation.

%\end{multicols}
%\begin{widetext}
% %%%%%%%%%%%%%%%%%%%%%%%%%%%%%%%%%%%%%%%%%%%%%%%%%%%%%%%%%%%%%%%%%%%%%%%%%%%%%%
% %%%%%%%%%%%%%%%%%%%%%%%%%%%%%%%%%%%%%%%%%%%%%%%%%%%%%%%%%%%%%%%%%%%%%%%%%%%%%%%
%\end{widetext}

\begin{figure}[t]
\includegraphics[trim={0cm 0cm 0cm 0cm},width=0.37\textwidth,angle=0,clip]{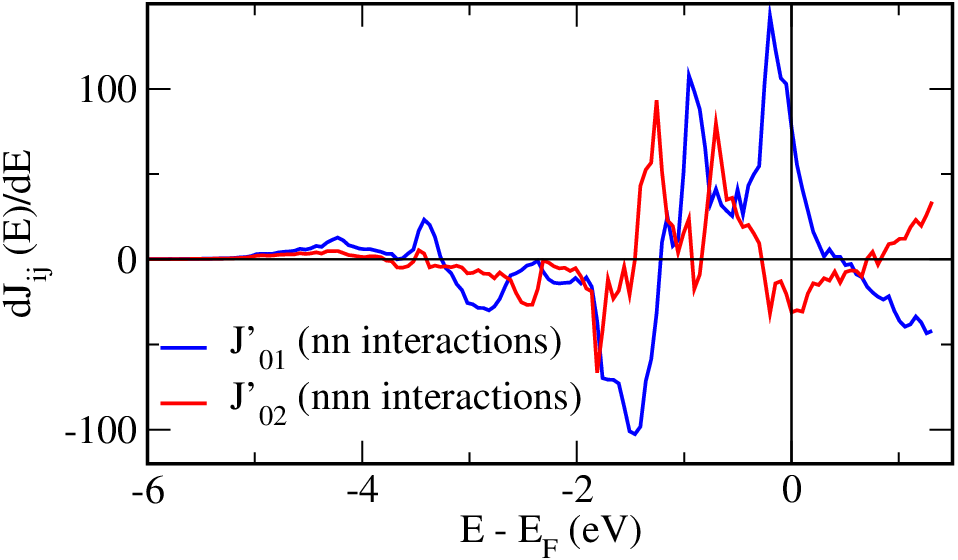}\,(a)
\includegraphics[trim={0cm 0cm 0cm 0cm},width=0.37\textwidth,angle=0,clip]{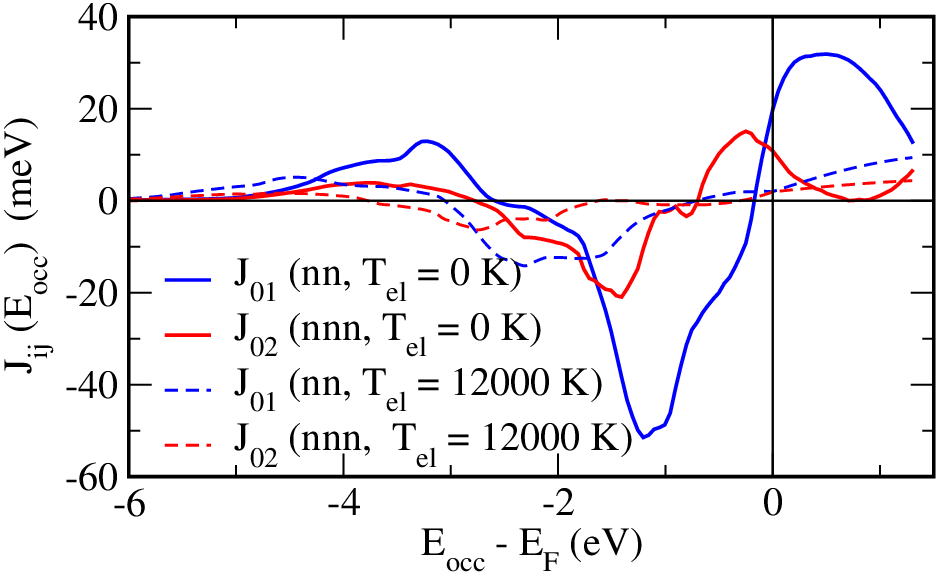}\,(b)
\caption{(a) The integrand in the second term of Eq.\ (\ref{Eq:NEW_Jij})
  for the interatomic nn and nnn exchange coupling parameters of bcc Fe
  in the ground state ($T_{\rm el} = 0$~K) as a function of energy; 
  (b) interatomic nn and nnn exchange coupling parameters,
  $J_{01}$ and $J_{02}$, respectively, represented as a function of
  occupation of the electronic states characterized by the threshold
  energy $E_{\rm occ}$ below or above the Fermi energy $E_{\rm F}$. Solid lines
  represent results obtained for the electronic temperature $T_{\rm el}
  = 0$~K, while dashed lines correspond to $T_{\rm el} = 12000$~K.
}
\label{fig:Fe_JXC_Eocc}
\end{figure}

 To get more insight into the behavior of the $J_{ij}$ parameters, 
  Fig.\ \ref{fig:Fe_JXC_Eocc}(a) shows the energy dependence of the
  integrand in the second term of Eq.\ (\ref{Eq:NEW_Jij}) which determines
  the interatomic nearest-neighbor (nn) and next-to-nearest-neighbor
  (nnn)  exchange coupling parameters $J_{01}$ and $J_{02}$,
  respectively, for bcc Fe in the ground state, i.e. $T = 0$~K. Note that the 
  first term in Eq.\ (\ref{Eq:NEW_Jij}) vanishes in this case because of
  the distribution function $f(E,T_{\rm el}=0) = \theta(E-E_F)$. 
  For both curves in Fig.\ \ref{fig:Fe_JXC_Eocc}(a),
  one can see a pronounced variation in the energy region  
  close to the Fermi energy. The corresponding integrals in Eq.\
  (\ref{Eq:NEW_Jij}), using the
  energy $E_{\rm occ}$ as an upper limit of integration, represent 
  the parameters $J_{01}(E_{\rm occ})$  and $J_{02}(E_{\rm occ})$ as a
  function of occupation of the electronic states. As can be seen in
  Fig.\ \ref{fig:Fe_JXC_Eocc}(b), these parameters exhibit a strong
  dependence on the occupation of the electronic states (see, e.g.,
  discussions in Ref.\ [\onlinecite{MPE20}]).
  A crucial consequence of the strong energy dependencies shown in Fig.\
  \ref{fig:Fe_JXC_Eocc}(a) occurs for the out-of-equilibrium electron
  subsystem with the laser-induced deoccupied states below $E_F$, and
  occupied states above $E_F$. In this case the main changes of the exchange
  parameters occur as a result of a convolution of the integrand plotted in
  Fig.\ \ref{fig:Fe_JXC_Eocc}(a) with the occupation function $f(E)$
  shown in  Fig.\ \ref{fig:Fe_TDDFT_JXC-t}. This can be seen in Fig.\
  \ref{fig:Fe_JXC_Eocc}(b), dashed line, representing the
  $J_{01}(E_{\rm occ},T)$  and $J_{02}(E_{\rm occ},T)$ parameters obtained using
  the occupation function $f(E,T_{\rm el}=12000 K)$ (see Fig.\ \ref{fig:Fe_JXC-Temper}).

% %%%%%%%%%%%%%%%%%%%%%%%%%%%%%%%%%%%%%%%%%%%%%%%%%%%%%%%%%%%%%%%%%%%%%%%%%%%%%%
% %%%%%%%%%%%%%%%%%%%%%%%%%%%%%%%%%%%%%%%%%%%%%%%%%%%%%%%%%%%%%%%%%%%%%%%%%%%%%%%

\subsection{Exchange coupling parameters under laser pulse and ultrafast demagnetization}

As a next step, we focus on the magnetic properties of a material
determined by laser induced changes of the exchange interactions, considering 
also bcc Fe as an example and making use of Eq.\ (\ref{Eq:NEW_Jij}). 
In this case we use as an input the exchange field $B_{\rm xc} =
\frac{1}{2}(V^\uparrow - V^\downarrow)$ and 
spin-resolved  occupation function $f_\sigma(E)$, delivered by TD-DFT
calculations performed with the Elk electronic structure code \cite{ELK},
for a laser pulse polarized along the $x$ axis, a fluence of 24
mJ/cm$^2$, a frequency of 413 THz and full width
at half maximum (FWHM) of 4.8 fs. 
%\end{multicols}
%\begin{widetext}
% %%%%%%%%%%%%%%%%%%%%%%%%%%%%%%%%%%%%%%%%%%%%%%%%%%%%%%%%%%%%%%%%%%%%%%%%%%%%%%
\begin{figure}[h]
\includegraphics[trim={0cm 0cm 0cm 0cm},width=0.37\textwidth,angle=0,clip]{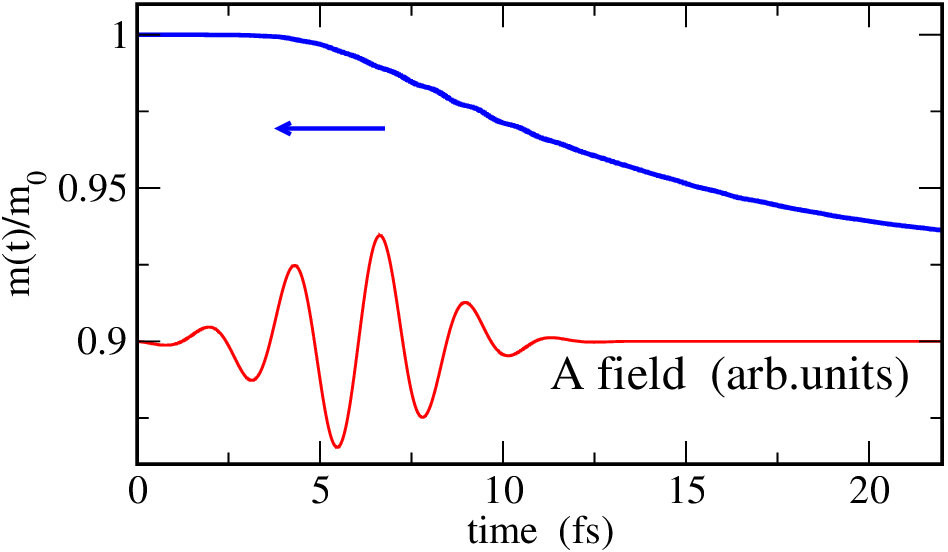}\,(a)
\includegraphics[trim={0cm 0cm 0cm 0cm},width=0.37\textwidth,angle=0,clip]{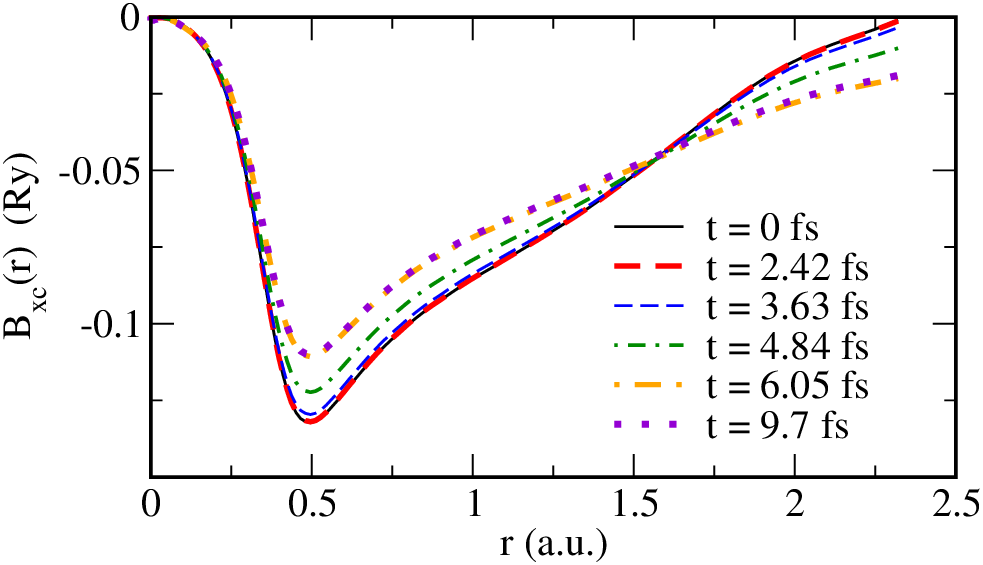}\,(b)
\caption{(a) Laser-pulse induced magnetization for bcc Fe. TD-DFT
  calculations using a laser pulse polarized along the $x$ axis, fluence 24
  mJ/cm$^2$,  FWHM of 4.8 fs. (b) Radial dependence of the effective
  exchange field $B_{\rm xc}(r) = \frac{1}{2}(V^{\uparrow} - V^{\downarrow})$ for Fe within the muffin-tin sphere,
  plotted for different time steps during the laser pulse.
}
\label{fig:Fe_TDDFT_m_t}
\end{figure}
% %%%%%%%%%%%%%%%%%%%%%%%%%%%%%%%%%%%%%%%%%%%%%%%%%%%%%%%%%%%%%%%%%%%%%%%%%%%%%%%
%\end{widetext}
%\begin{multicols}
The TD-DFT calculations have been performed for the time interval up to 22 fs.
As one can see in Fig. \ref{fig:Fe_TDDFT_m_t} (a), the magnetization
decreases during the laser pulse by $\sim 4\%$ due to the
SOC-induced spin-flip transitions. It continues to decrease also after
the laser pulse down to $\sim 8\%$,  
 due to the relaxation within the electronic subsystem
  accompanied by SOC-induced spin-flip transitions
  between the 'up' and 'down' spin channels characterized by different 
  out-of-equilibrium occupation functions $f_{\sigma}(E,t)$.

% %%%%%%%%%%%%%%%%%%%%%%%%%%%%%%%%%%%%%%%%%%%%%%%%%%%%%%%%%%%%%%%%%%%%%%%%%%%%%%
\begin{figure*}[t]
\includegraphics[trim={0cm 0cm 0cm 0cm},width=0.17\textwidth,angle=0,clip]{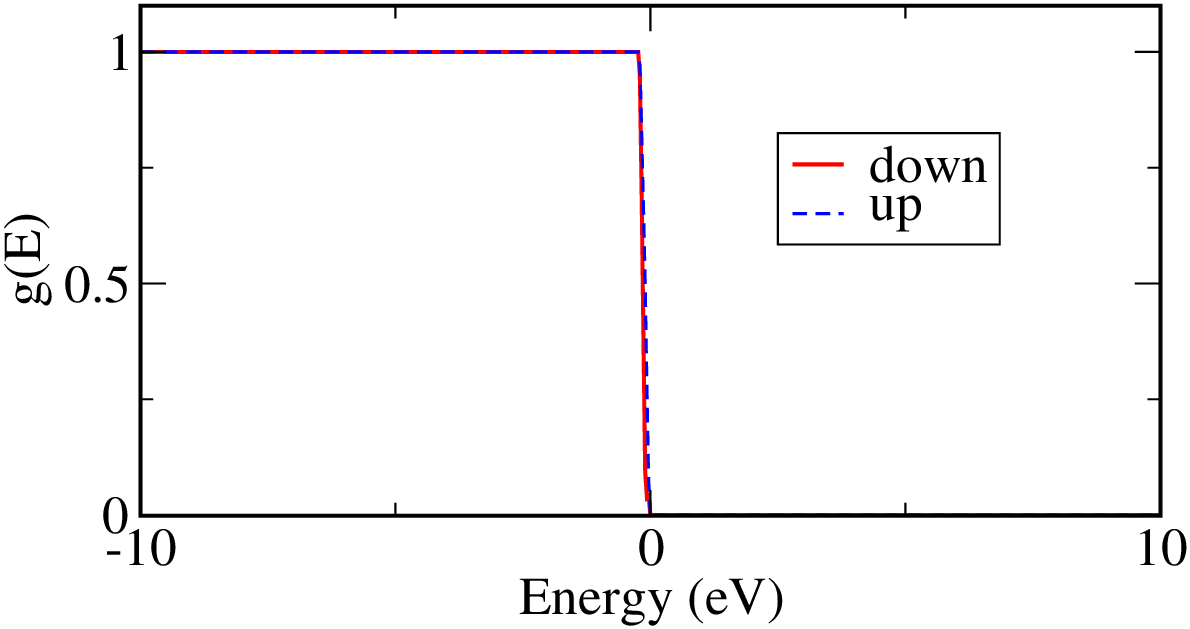} \,\,\,\,\,\,\,\,
\includegraphics[trim={0cm 0cm 0cm 0cm},width=0.17\textwidth,angle=0,clip]{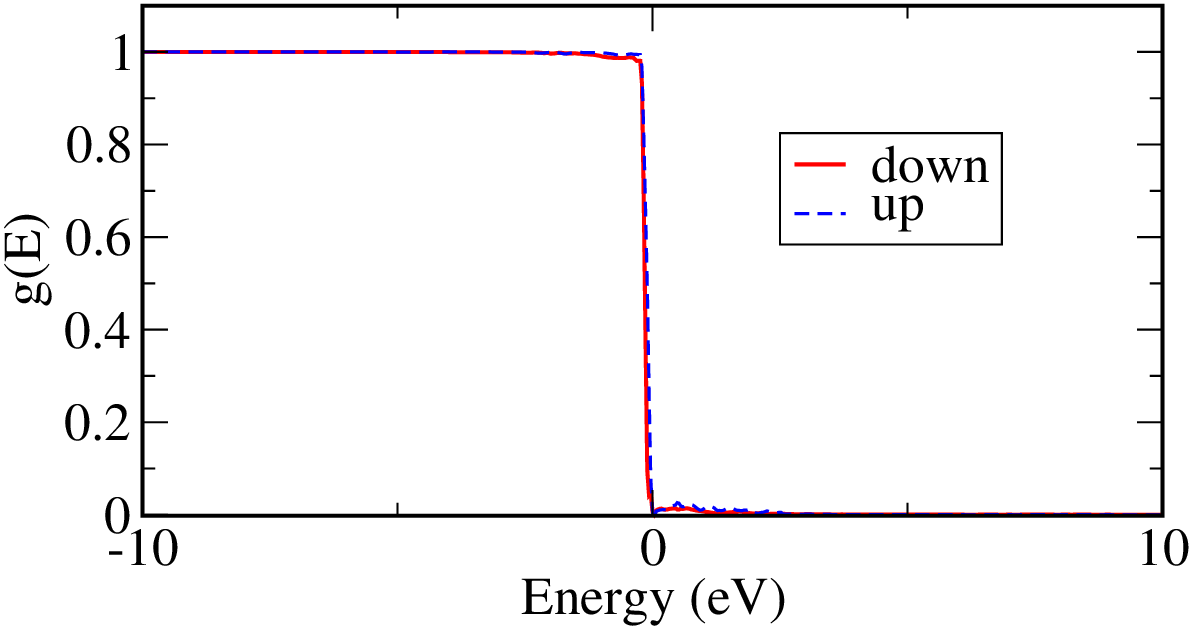}               \,\,\,\,\,\,\,\,  
\includegraphics[trim={0cm 0cm 0cm 0cm},width=0.17\textwidth,angle=0,clip]{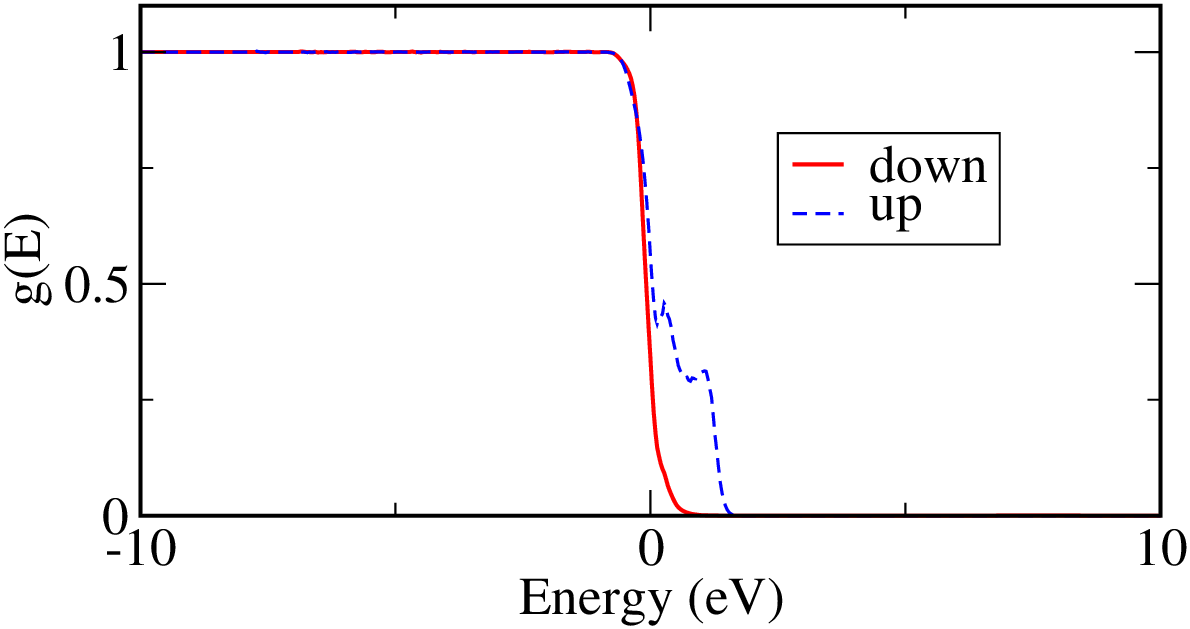} \,\,\,\,\,\,\,\,
\includegraphics[trim={0cm 0cm 0cm 0cm},width=0.17\textwidth,angle=0,clip]{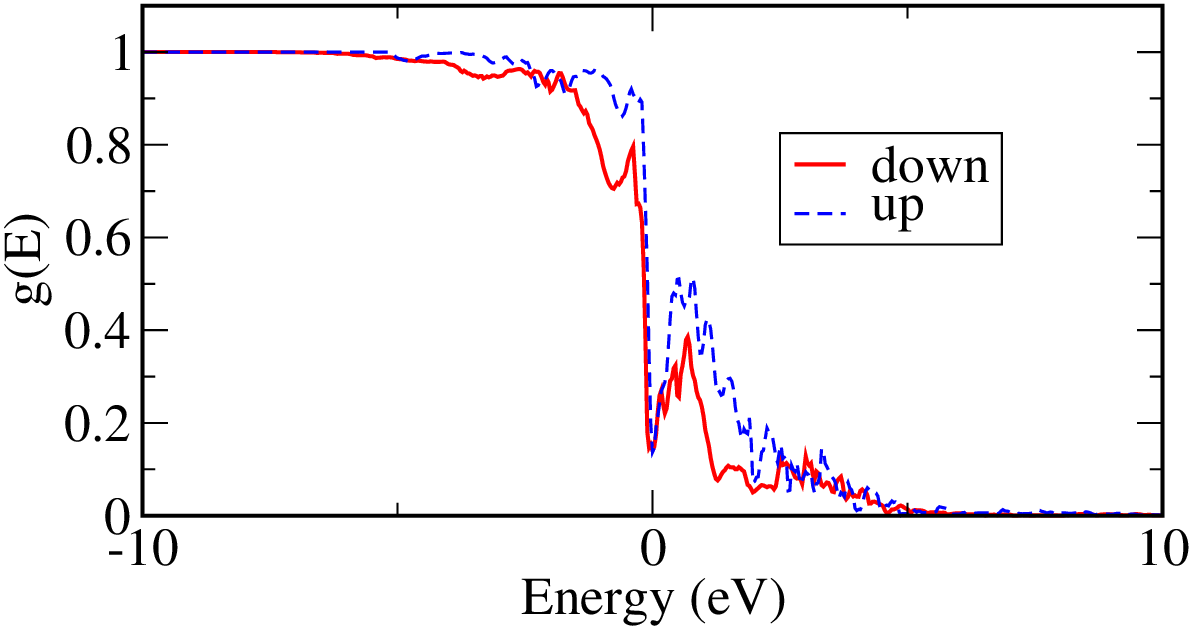}    \,\,\,\,\,\,\,\,             
\includegraphics[trim={0cm 0cm 0cm 0cm},width=0.17\textwidth,angle=0,clip]{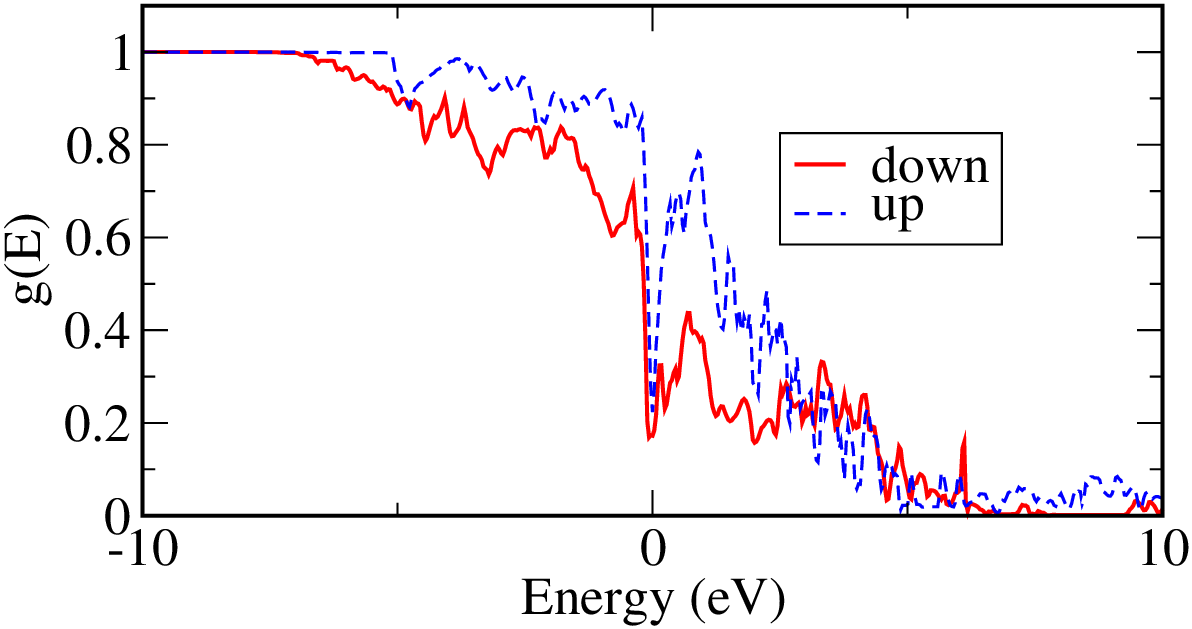} \,\,\,\,\,\,\,\,
\includegraphics[trim={0cm 0cm 0cm 0cm},width=0.17\textwidth,angle=0,clip]{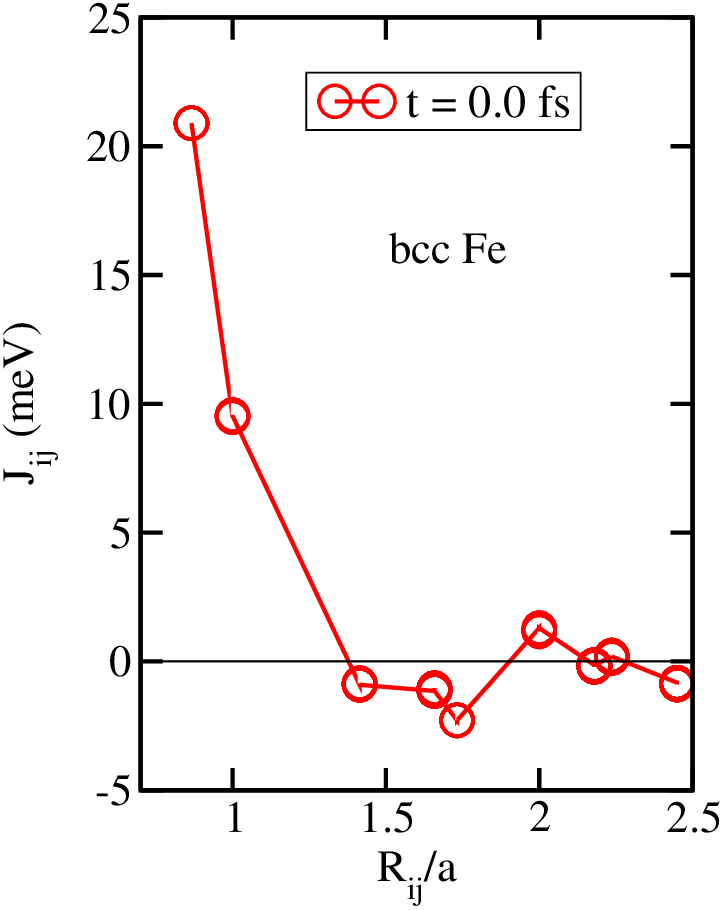}\,\,\,\,\,\,\,\,                 
\includegraphics[trim={0cm 0cm 0cm 0cm},width=0.17\textwidth,angle=0,clip]{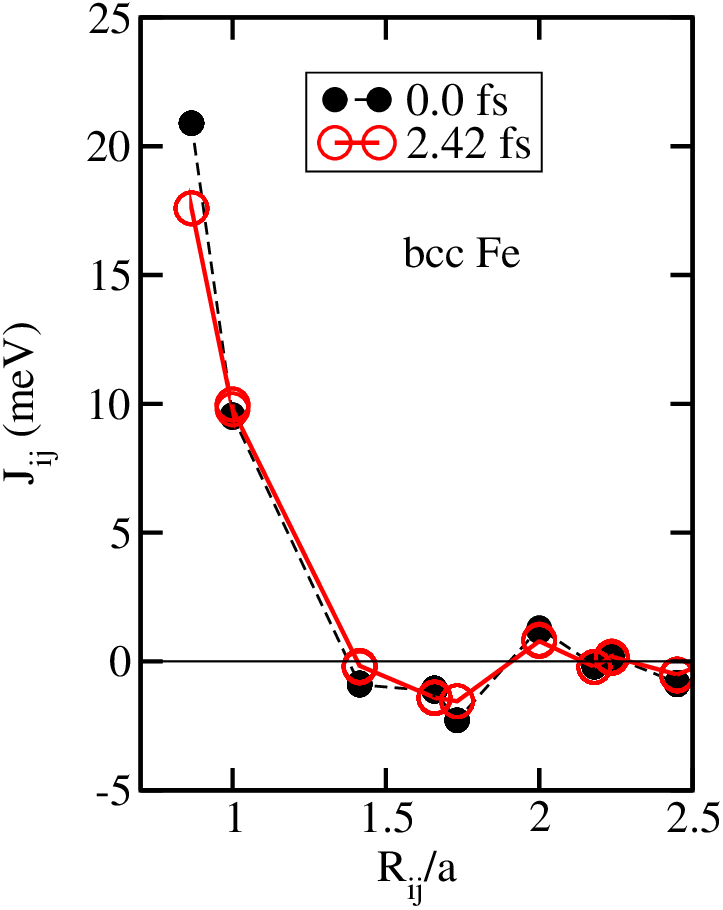} \,\,\,\,\,\,\,\,
\includegraphics[trim={0cm 0cm 0cm 0cm},width=0.17\textwidth,angle=0,clip]{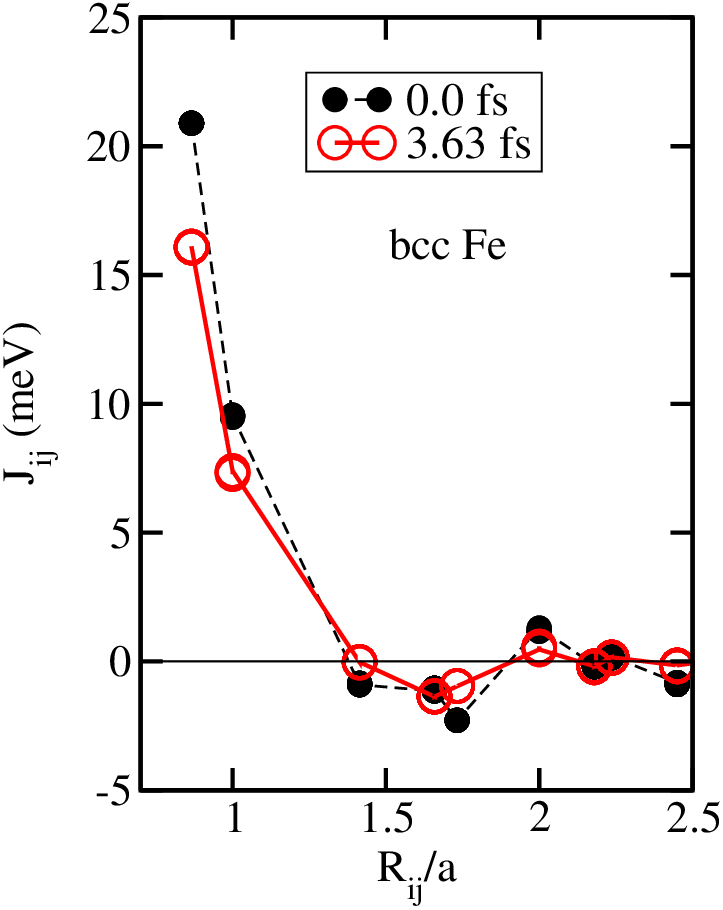}\,\,\,\,\,\,\,\,
\includegraphics[trim={0cm 0cm 0cm 0cm},width=0.17\textwidth,angle=0,clip]{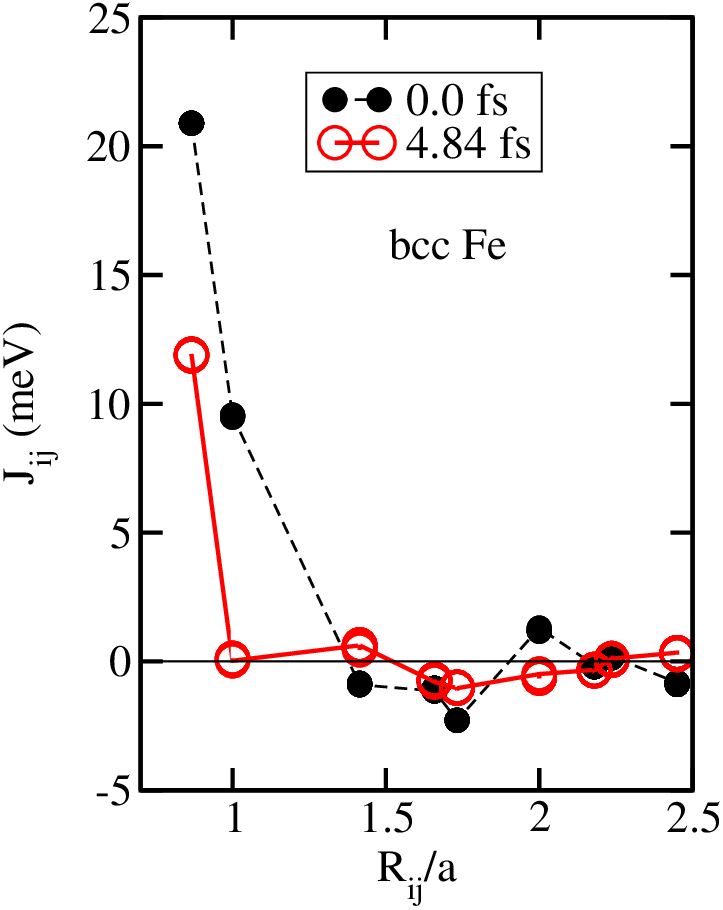}\,\,\,\,\,\,\,\,
\includegraphics[trim={0cm 0cm 0cm 0cm},width=0.17\textwidth,angle=0,clip]{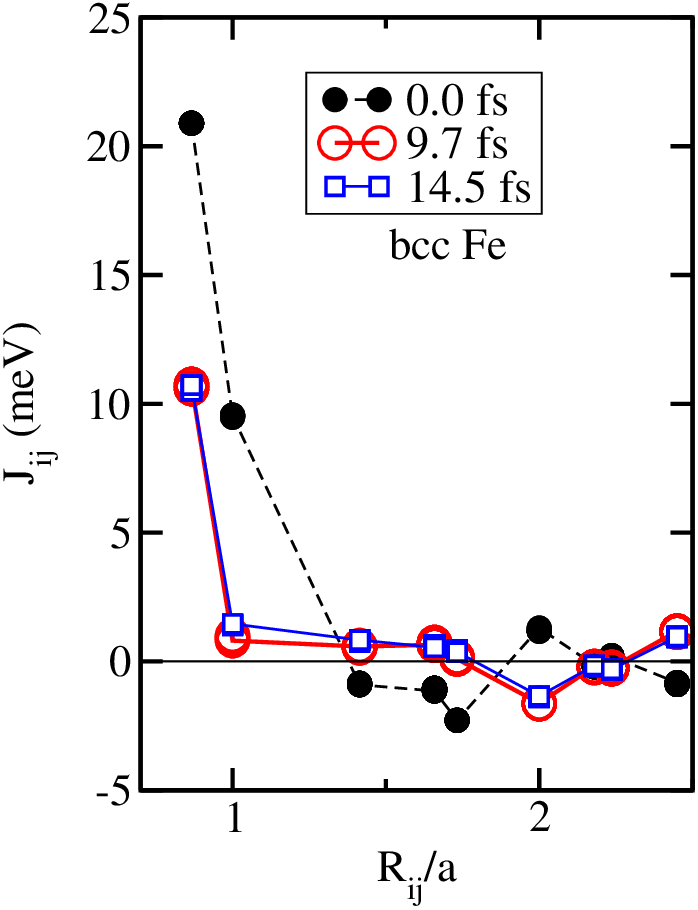}\,\,\,\,\,\,\,\,
\includegraphics[trim={0cm 0cm 0cm 0cm},width=0.17\textwidth,angle=0,clip]{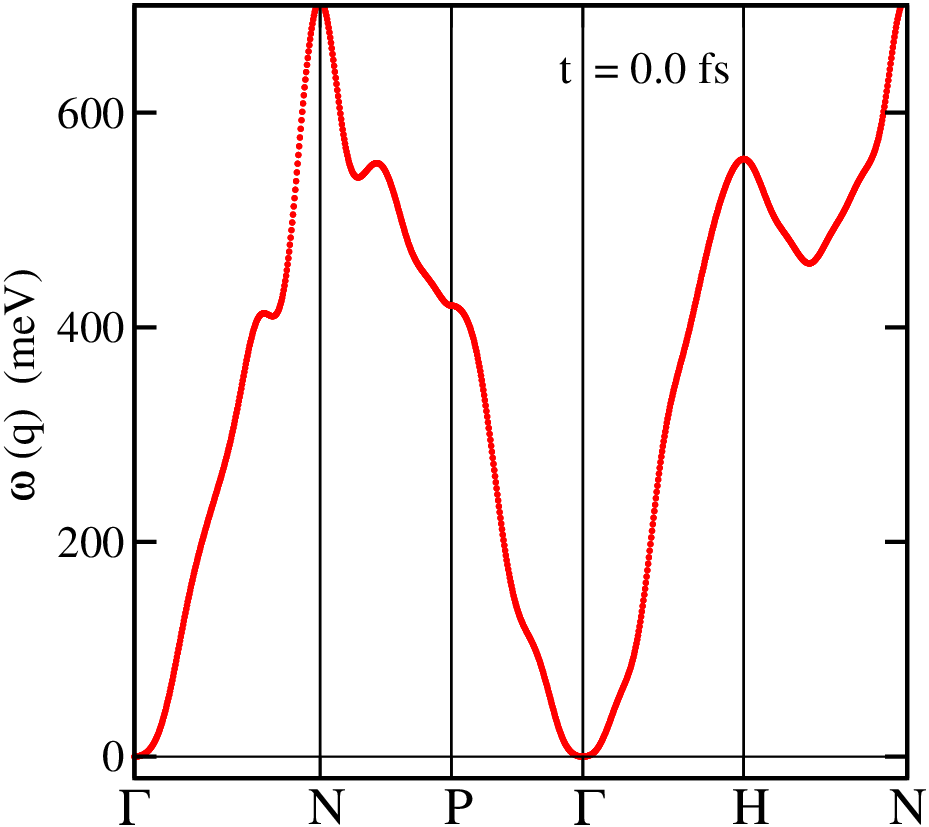}(a)
\includegraphics[trim={0cm 0cm 0cm 0cm},width=0.17\textwidth,angle=0,clip]{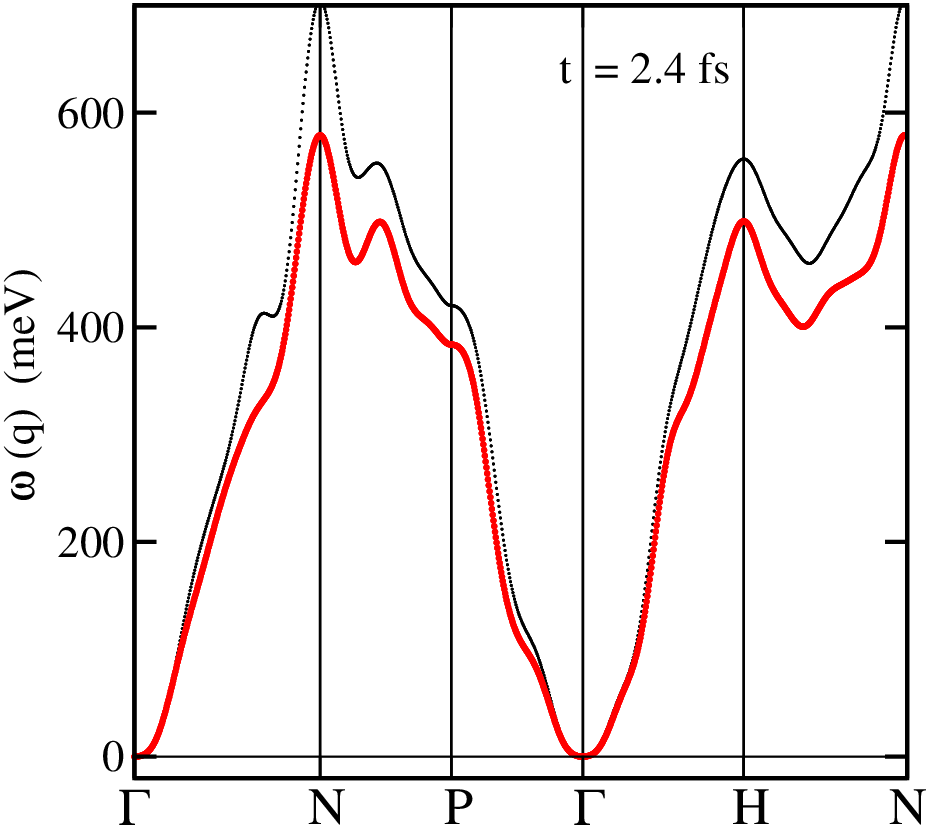}(b)
\includegraphics[trim={0cm 0cm 0cm 0cm},width=0.17\textwidth,angle=0,clip]{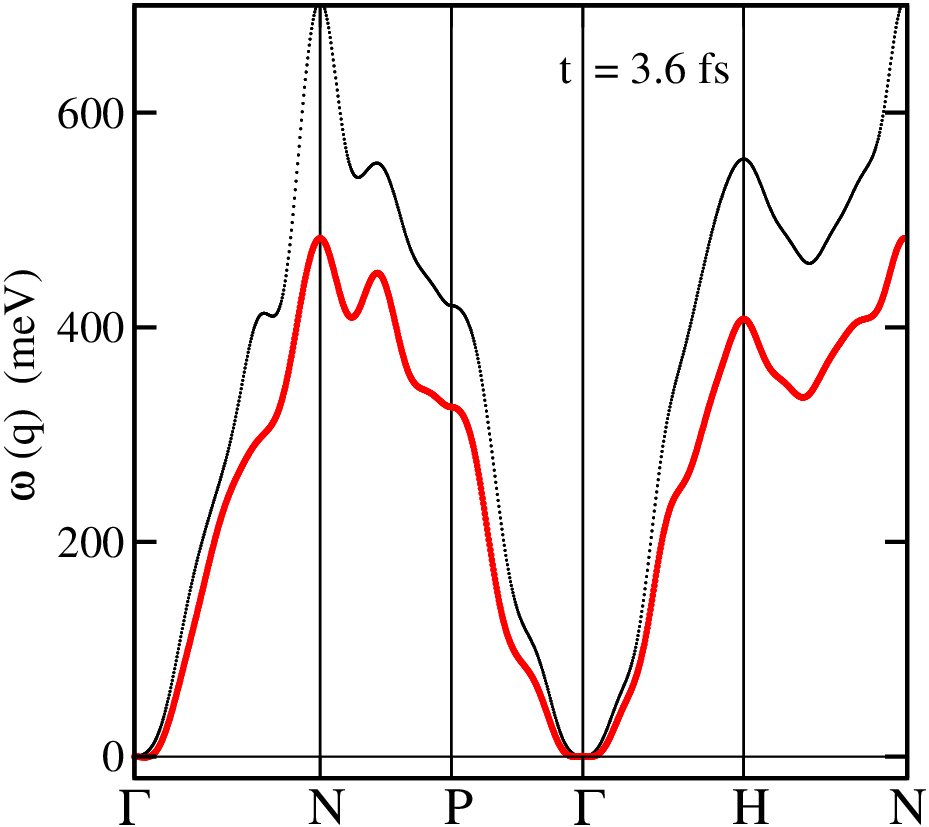}(c)
\includegraphics[trim={0cm 0cm 0cm 0cm},width=0.17\textwidth,angle=0,clip]{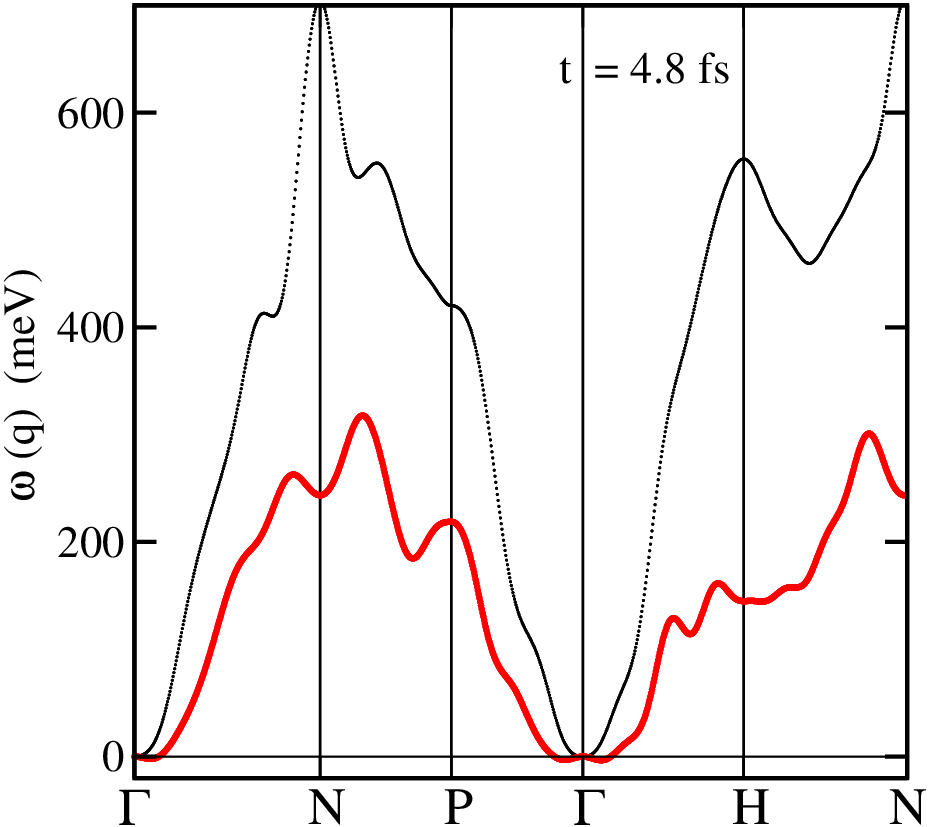}(d)
\includegraphics[trim={0cm 0cm 0cm 0cm},width=0.17\textwidth,angle=0,clip]{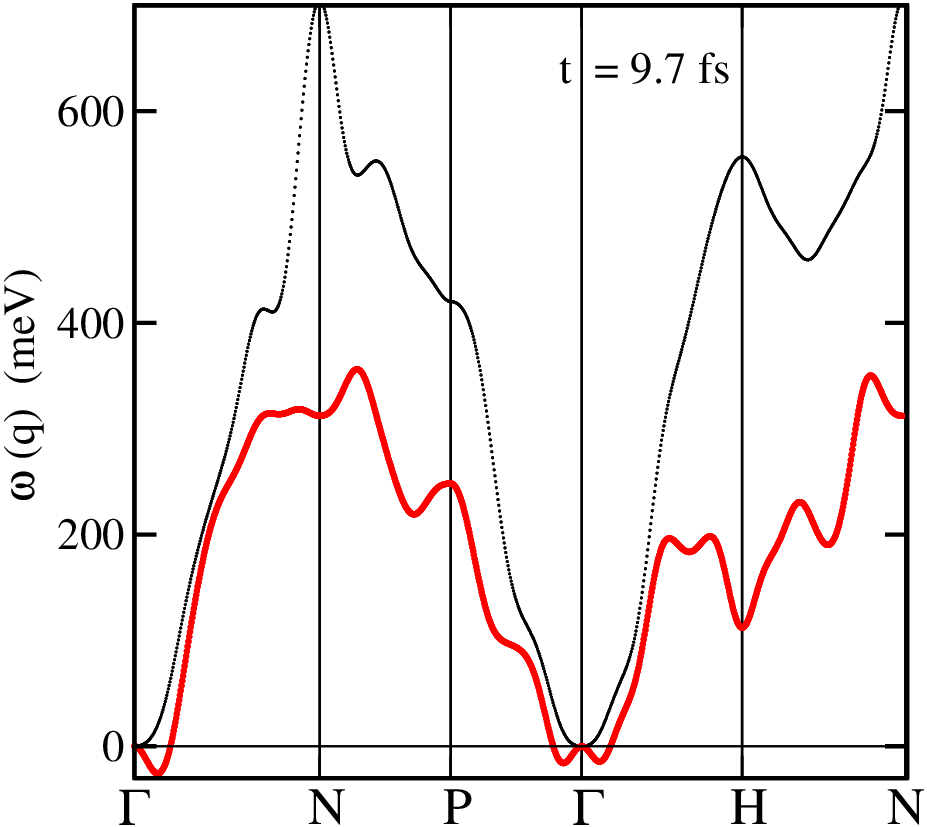}(e)
\caption{(Top) spin-dependent electronic occupation function $f_{\sigma}(E,t)$ for five time
  steps during the laser pulse, calculated using the Elk TD-DFT code, with
  the laser pules parameters: FWHM = 4.8 fs, pulse fluence of 24
  mJ/cm$^2$, light frequency 413 THz. (Middle) The exchange
  coupling parameters $J_{ij}$ (open symbols) calculated for several time steps with
  the SPR-KKR band structure code using corresponding occupation
  functions $f_{\sigma}(E,t)$ (top panel) and effective exchange field
  $B_{\rm xc}$ obtained
  within TD-DFT calculations. Full symbols represent the results for the
  ground state parameters. (Bottom) Spin-wave dispersion relations
  calculated for different time steps (red symbols) using the exchange
  parameters shown in the middle panel. Black symbols represent the
  results for the results for the ground state.
}
\label{fig:Fe_TDDFT_JXC-t}
\end{figure*}
% %%%%%%%%%%%%%%%%%%%%%%%%%%%%%%%%%%%%%%%%%%%%%%%%%%%%%%%%%%%%%%%%%%%%%%%%%%%%%%%
%\end{widetext}
%\begin{multicols}

The decrease of the Fe magnetic moment is a consequence of the modified spin
density due to the re-population of the
electronic states, induced by the laser pulse. This in particular results in a
change of the effective exchange field $B_{\rm xc}(r)$ shown 
in Fig. \ref{fig:Fe_TDDFT_m_t} (b),
plotted within the muffin-tin sphere for different time steps.
One can clearly see a decrease of the magnitude of the B-field during
the laser pulse, while the changes are almost invisible when the pulse
is finished. 

The laser-induced spin-dependent occupation function $f_{\sigma}(E,t)$ plotted in
Fig.\ \ref{fig:Fe_TDDFT_JXC-t}, top panel, is represented for five time steps
during the laser pulse. 
One can see a considerable depopulation  of the electronic states below
$E_F$ due to the optical excitation to high-energy
states. Moreover, in contrast to finite-temperature considerations discussed above,
the occupation functions are different for the majority- and minority-spin
states, and the difference increases during the laser pulse.

The quantities $B_{\rm xc}(r)$ and $f_{\sigma}(E,t)$ have been used in a
quasi-stroboscopic way for   
calculations of the Green function varying in time under the influence
of the laser pulse, to be used later on for calculation of time-dependent
exchange coupling parameters. The calculations have been performed by
means of the SPR-KKR band structure method using $B_{\rm xc}(r)$ 
at every time step to calculate the corresponding scattering path
operator and the perturbation potential $\langle \Delta V
\rangle_i^\alpha$.
The exchange coupling parameters, $J_{ij}$, modified under the
impact of the laser pulse are given in the middle panel of Fig.\
\ref{fig:Fe_TDDFT_JXC-t}. As one would expect, the $J_{ij}$ parameters decrease with 
time during the laser pulse. The most pronounced changes concern the
nearest-neighbor and next-nearest-neighbor parameters.
 After the pulse, the exchange parameters remain practically unchanged
  (see Fig. \ref{fig:Fe_TDDFT_JXC-t}(e), middle panel). 
  
%As one can see in Fig. \ref{fig:Fe_TDDFT_JXC-t}(e) (middle panel), after
%the pulse the exchange parameters are practically unchanged.

% To visualize the consequences to be expected from such a modification of
% the exchange interactions, one can consider the impact of these changes
% on the magnetic properties of the system when it is seen to be
% electronically in a frozen state at every time step. 
% As Fig.\  \ref{fig:Fe_TDDFT_JXC-t} demonstrates (bottom panel), the
% change in $J_{ij}$ leads to a softening of the magnon spectrum
% reflecting the fact that the magnetization gets less robust.

\begin{figure}[t]
\includegraphics[trim={0cm 0cm 0cm 0cm},width=0.37\textwidth,angle=0,clip]{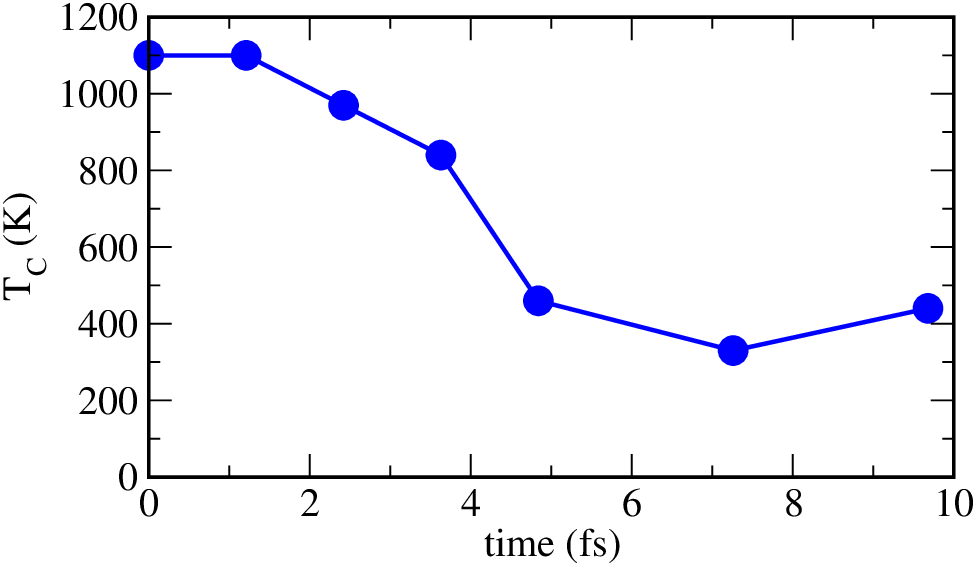}\,(a)
\includegraphics[trim={0cm 0cm 0cm 0cm},width=0.37\textwidth,angle=0,clip]{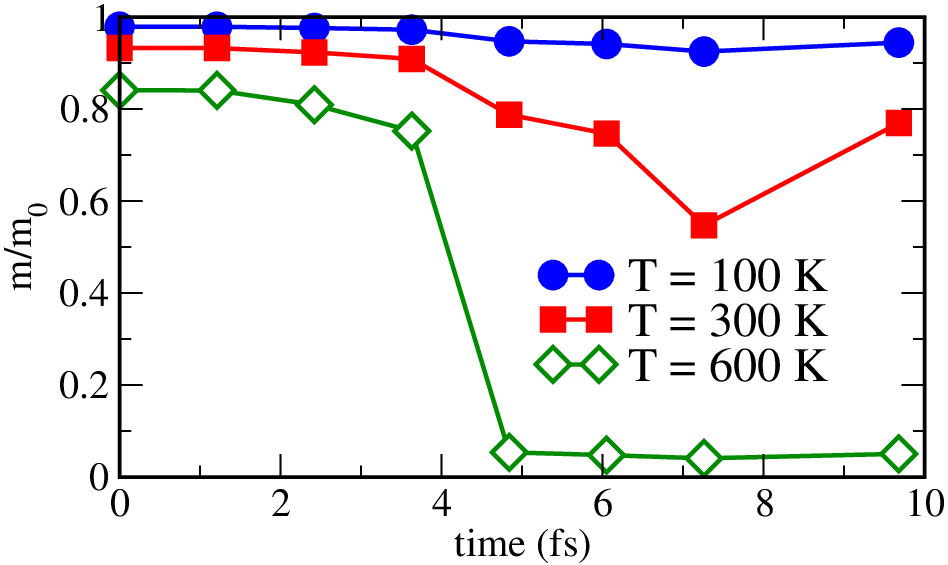}\,(b)
\caption{(a) Curie temperature for bcc Fe as a function of time calculated
  using the exchange coupling parameters changing in time due to
  the laser pulse induced evolution of the electronic structure. (b)
  Time-dependent evolution of the magnetization for different
  temperature of the spin subsystem taken unchanged at the time interval
  shown in the plot. Even neglecting the energy transfer from the
  electronic subsystem in the ultrafast regime, the temperature can be
  seen as an initial temperature of the sample being in thermodynamic equilibrium.
}
\label{fig:Fe_TDDFT_TC_t}
\end{figure}

To illustrate, which impact on the magnetic properties can be expected
due to  such a modification of the exchange interactions, one can
consider the system at every time step as one being electronically
in a frozen state. 
As can be seen in Fig.\ \ref{fig:Fe_TDDFT_JXC-t} (bottom), 
the modification of the exchange interactions shown in  Fig.\
\ref{fig:Fe_TDDFT_JXC-t} (middle) leads to a softening of the magnon
spectrum, reflecting the fact that the magnetization gets less robust.
In line with this, we find a distinct decrease of the effective Curie temperature
during the absorption of the laser pulse (see
Fig. \ref{fig:Fe_TDDFT_TC_t}(a)), which was obtained from 
Monte Carlo simulations using the fixed exchange coupling parameters.
  Furthermore, considering the initial temperature for the spin subsystem,
  $T_{\rm Heis}$, to be finite and fixed during the laser pulse,
  which seems to be a reasonable approximation for the ultrashort time scale, one can
  check the time dependent behavior of the net magnetization using different
  exchange parameters calculated for different time steps.
  Corresponding results for the magnetization evolution
  obtained making use of the parameters shown in Fig. \ref{fig:Fe_TDDFT_JXC-t}
  are plotted in  Fig. \ref{fig:Fe_TDDFT_TC_t}(b) for three different
  temperatures  $T_{\rm Heis}$, demonstrating a strong magnetization decrease during the laser
  pulse which can lead even to the break of the magnetic order if $T_{\rm Heis}$
  is high enough.
  Obviously, these conclusions on the magnetic properties have essentially
  semi-qualitative character because of the approximations discussed
  above. Nevertheless, we would like to stress that they are based on
  the electronic structure with its time evolution coherently treated
  within the TD-DFT formalism.   

%\begin{widetext}
% %%%%%%%%%%%%%%%%%%%%%%%%%%%%%%%%%%%%%%%%%%%%%%%%%%%%%%%%%%%%%%%%%%%%%%%%%%%%%%
% %%%%%%%%%%%%%%%%%%%%%%%%%%%%%%%%%%%%%%%%%%%%%%%%%%%%%%%%%%%%%%%%%%%%%%%%%%%%%%%
%\end{widetext}
%\begin{multicols}

\begin{figure}[t]
\includegraphics[trim={0cm 0cm 0cm 0cm},width=0.4\textwidth,angle=0,clip]{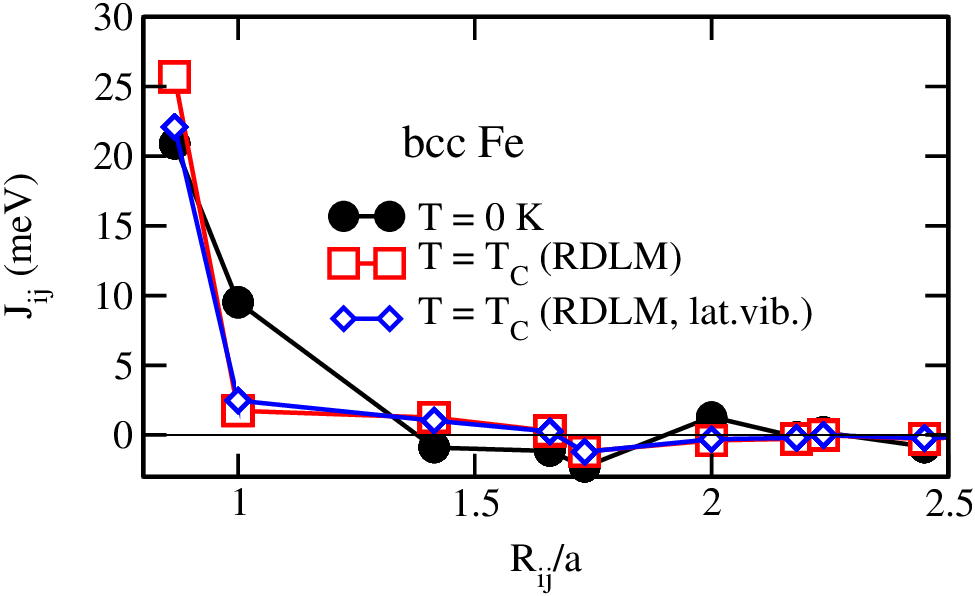}\,
\caption{Exchange coupling parameters $J_{ij}$ for bcc Fe calculated for
  the FM reference state (full circles), for the magnetically disordered
  states treated within the RDLM approch (opened squres) and accounting
  both magnetic disorder and atomic diaplacements at $T = T_C$.
  The amplitude of the lattice vibrations at given temperature is estimated
  using the Debye  model \cite{EMC+15,MPE20}. 
}
\label{fig:Fe_RDLM_Jij}
\end{figure}

  In a more realistic scenario, $T_{\rm Heis}$ is non-zero and can change also
  during and after the laser pulse, that happens however on a longer time
  scale. This implies a magnetic disorder in the system varying
  in time until the full relaxation of the excited system.
  As the exchange parameters depend also on the magnetic configuration,
  this situation can also be taken into account considering the laser induced
  demagnetization and relaxation processes.
  Such investigations have been performed recently combining
  first principles calculations and spin dynamics
  simulations\cite{DHSN17}, where the finite temperature effect after
  the laser pulse has been taken into account for the electron and spin
  subsystems. However, in that work, both effects were
  small because of a weak laser power used in the calculations 
  (for comparison, the maximal effective temperature of the electron gas
  considered in this work was about 1000 K, far below the
  effective temperature $\sim 10^4$ K estimated for the laser pulse used
  in the present case.)  
  Furthermore, the thermalization of the spin and lattice subsystems at
  later times results in an increase of the magnetic disorder as
  well as increase of the amplitude of atomic displacements.
  One can compare the impact of these factors on the exchange coupling
  with the impact of the laser-induced strong excitation of the electron 
  subsystem. For this we consider the limiting case of $T = T_C$ with
  thermally induced magnetic disorder treated here making use of the
  relativistic disordered local moment (RDLM) theory
  \cite{SSB+06,EMC+15}.  
  Corresponding exchange coupling parameters $J_{ij}$ are
  plotted by squares in Fig. \ref{fig:Fe_RDLM_Jij}. As one can see, the
  nnn parameters decrease significantly when compared to the FM system
  ($T=0$K), however at the same time one can see an increase of the nn
  parameters. Obviously, this effect is very different when compared to
  the effect caused by the laser-excited out-of-equilibrium electron
  subsystem. 
% %%%%%%%%%%%%%%%%%%%%%%%%%%%%%%%%%%%%%%%%%%%%%%%%%%%%%%%%%%%%%%%%%%%%%%%%%%%%%%
% %%%%%%%%%%%%%%%%%%%%%%%%%%%%%%%%%%%%%%%%%%%%%%%%%%%%%%%%%%%%%%%%%%%%%%%%%%%%%%%
  The lattice vibrations at this temperature \cite{MPE20} also lead to some changes
  of $J_{ij}$,  which however mainly concern the nn parameters.

\subsection{Spin-lattice coupling parameters}
    
 Finally we will shortly discuss another type of interaction,
  namely the spin-lattice interactions responsible for the energy and
  angular momentum transfer between the magnon and phonon subsystems.
These interactions are discussed, in particular, for a long time scale
when considering the mechanisms responsible for the magnetic relaxation
\cite{MPL+22}. 

 The impact of the lattice vibrations on the exchange coupling
 parameters is demonstrated in Fig. \ref{fig:Fe_RDLM_Jij}.
  Discussing this in terms of the spin-lattice interaction, it was
 shown \cite{MLPE23} that it can be associated with the SLC parameters
 of second order with respect to displacements, $J^{\alpha\alpha,\mu\nu}_{ij,kl}$,
 leading to corrections of the exchange coupling $\Delta J_{ij} \sim
 \langle u^2\rangle_T$. Note that the amplitude of the atomic
   displacement is directly connected to the lattice temperature $T_{\rm
     lat}$, and can be reasonably well estimated for a given temperature $T_{\rm
     lat}$ within the Debye model approach. 
 A more pronounced effect is expected due to the SLC parameters linear
 w.r.t. the atomic displacements. These parameters characterize the rate of 
 spin angular momentum transfer due to the torque on the magnetic moments, $ \vec{\cal
   T}_i = \hat{e}_i \times \vec{H}_{\rm eff}$, caused by the effective field
 induced by the displacements  $u_k^\mu$ of the atoms on sites $k$,
 or alternatively, due to the mechanical torque on the atom $k$,
$\vec{\mathfrak{T}}^{ph}_k = \vec{u}_k \times 
\vec{\cal F}_k$, created by the forces induced by spin tiltings
$\hat{e}^\alpha_i$  on sites $i$ \cite{MPL+22,MLPE23,RSBD20}. 

Similar to the exchange coupling parameters, the SLC parameters are
determined by the electronic structure and one can expect their changes 
caused by the laser induced strong modification of the electron subsystem.
We will focus here only on the so-called DMI-like SLC parameters
$\vec{\cal D}^{\mu}_{ij,k}$ \cite{MPL+22,MLPE23}, characterizing the
change of the DMI parameters $\Delta \vec{D}_{ij} =  \sum_{k,\mu}\vec{\cal
  D}^{\mu}_{ij,k}(u_k^\mu - u_i^\mu)$  caused by atomic displacements
$\vec{u}_k$ (note that similar arguments hold also when considering the
properties of the MCA-like SLC parameters,
${\cal K}^{\alpha z,\mu}_{i,k}$ characterizing local changes of the MCA  parameters
$\Delta {K}^{\alpha z}_{i} = \sum_{k,\mu}{\cal K}^{\alpha
  z,\mu}_{i,k}(u_k^\mu - u_i^\mu)$).

\begin{figure}[t]
\includegraphics[trim={0cm 0cm 0cm 0cm},width=0.37\textwidth,angle=0,clip]{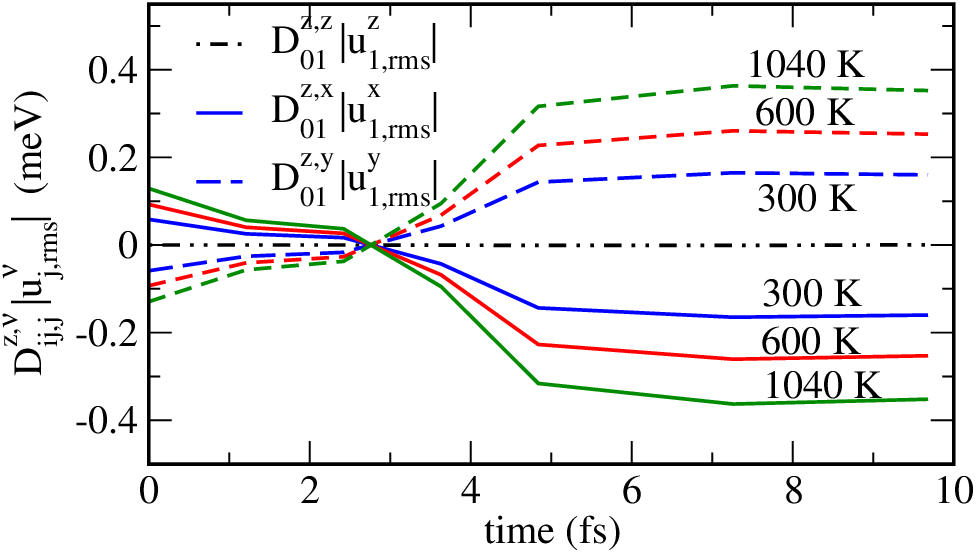}\,(a)
\includegraphics[trim={0cm 0cm 0cm 0cm},width=0.37\textwidth,angle=0,clip]{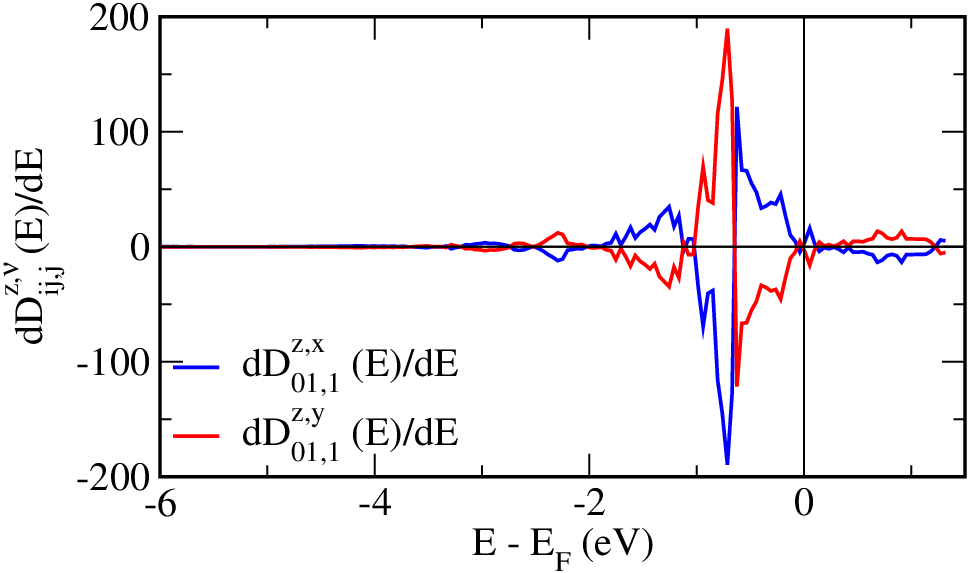}\,(b)
\includegraphics[trim={0cm 0cm 0cm 0cm},width=0.37\textwidth,angle=0,clip]{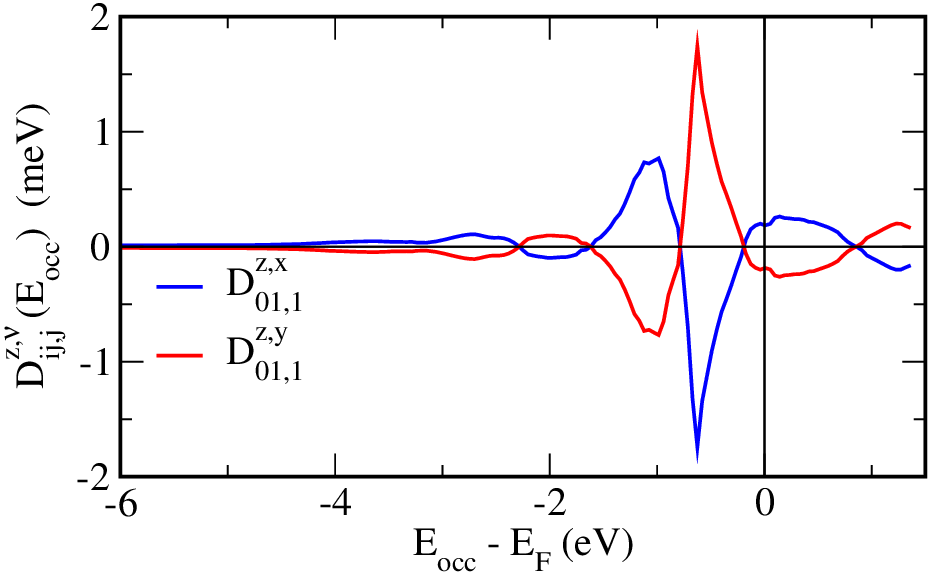}\,(c)
\caption{(a) DMI-like spin-lattice coupling parameters as a function of time, represented by a
  contribution to the DMI parameters due to atomic displacement, ${\cal
    D}^{z,\mu}_{01,1}(u_j^\mu - u_i^\mu)$ for $\mu = x,y,z$
  characterizing the directions of displacements of the atom in the
  position $\vec{R}_{01} =  (-0.5,-0.5,-0.5)a_{\rm lat}$. The results are
  presented for three different temperatures of the crystal lattice, 300
  K, 600 K and 1040 K ($T_C$), described by the displacement amplitude
  given by rms displacement $u_D(T)$ estimated within the Debye model,
  i.e. $0.18$, $0.28$ and $0.39$ a.u., respectively. (b) The integrands
  for the components of the DMI-like spin-lattice coupling parameters,
  ${\cal D}^{z,x}_{01,1}$ and  ${\cal D}^{z,y}_{01,1}$, as a
  function of energy.
  (c) The components
  of the DMI-like spin-lattice coupling parameters,
  ${\cal D}^{z,x}_{01,1}$ and  ${\cal D}^{z,y}_{01,1}$, represented as a
  function of occupation of the electronic states characterized by the
  threshold energy $E_{\rm occ}$ below or above the Fermi energy $E_{\rm F}$. 
}
\label{fig:Fe_SLC_time}
\end{figure}

As an example, Fig.\ \ref{fig:Fe_SLC_time} shows the parameters ${\cal
  D}^{z,\mu}_{ij,j}\, u_D(T)$ represented as a function of time for three
different temperatures of the crystal lattice $T_{\rm
     lat}$: 300 K, 600 K and 1040 K.
The lattice temperature is assumed to be constant during the full time range.
One can see, that the SLC parameters (and corresponding contributions to
$D^z_{ij}$) are of the order of $0.1$ meV in the equilibrium, and change
in amplitude and sign during the laser pulse reaching a value up to $\sim 0.4$ meV
(depending on the lattice temperature).
Similar arguments as in the case of the exchange coupling parameters may
be used also to explain the time dependent change of the ${\cal
  D}^{z,\mu}_{ij,j}$ parameters. This is a consequence the strong energy
dependent changes of the integrand (see Fig.\ \ref{fig:Fe_SLC_time}(b))
in the expression for the SLC parameter (see Refs.\
[\onlinecite{MPL+22,MLPE23}]), leading in particular to   
a strong dependence of the SLC on the occupation of the electron states shown in
Fig.\ \ref{fig:Fe_SLC_time}(c).

% %%%%%%%%%%%%%%%%%%%%%%%%%%%%%%%%%%%%%%%%%%%%%%%%%%%%%%%%%%%%%%%%%%%%%%%%%%%%%%
% %%%%%%%%%%%%%%%%%%%%%%%%%%%%%%%%%%%%%%%%%%%%%%%%%%%%%%%%%%%%%%%%%%%%%%%%%%%%%%%

\section{Summary}

 In summary, we find a distinct impact of the optically
overheated electron subsystem on the inter-atomic exchange coupling
parameters. While a simplified scheme is used to demonstrate this
effect, the calculations of the exchange parameters rely on
the electronic structure calculated using the TD-DFT method
which ensure an accurate description of the electronic structure
modifications under the influence of an ultrafast laser pulse.
We have demonstrated a crucial role of the laser-induced deoccupation of
the electronic states varying in time and dependent on the power of the laser
pulse, leading to a strong weakening of the exchange parameters.
Similarly, the laser-induced changes of the spin-lattice parameters
can influence the efficiency of magnon generation and in turn the
rate of demagnetization.
At a longer time scale the laser-induced changes of the exchange
parameters as well as spin-lattice parameters can influence the
magnetization relaxation process and should correct 
the results obtained so far using equilibrium parameters.

%\section{Acknowledgments}

%\clearpage

\section{Appendix}

\appendix

\section{Exchange coupling parameters out of equilibrium: impact of
  non-equilibrium occupation }\label{Appendix1}

Considering the FM state as a reference state, the deviation of some
magnetic moments from the collinear FM state is described by the perturbation
potential

\begin{eqnarray}
\Delta V(\vec{r}) & = & \sum_i \delta v(\vec{r}-\vec{R}_i) \nonumber \\ 
  &=& -\sum_i \beta (\vec{\sigma} \cdot \delta \hat{m}_i) B_i^{xc}(r) \; .
\label{EQ:perturb}
\end{eqnarray}

In the presence of a perturbation, the Green function  is modified as
follows

\begin{widetext}
  
\begin{eqnarray*}
G & = & G_0 + G_0 \Delta V G_0 +  G_0 \Delta V G_0 \delta V G_0 + ...  =  G_0 + \Delta G \; .
\end{eqnarray*}

Accounting only for single particle contribution to the free energy,
omitting the $\vec{r}$ arguments, the perturbation results in a free energy change 

\begin{eqnarray}
\Delta F & = & \left(-\frac{1}{\pi}\right) \mbox{Im}\mbox{Trace} \int^{\infty}_{E_{\rm bottom}} dE f(E) (E - E_F) \Delta G(E)  \; .
\end{eqnarray}

Keeping the second-order free energy term, one can use the sum rule for the Green functions $\frac{d G(E)}{dE} =
-G(E)G(E)$ to transform this expression and taking into account the
property of invariance of the trace under cyclic permutations of matrices, the expression above can be
transformed as follows 

\begin{eqnarray}
  {\cal F}^{(2)} & = &  \left(-\frac{1}{\pi}\right)
\mbox{Im}\mbox{Trace} \int^{\infty}_{E_{\rm bottom}} dE\, f(E)\, (E - E_F)\,\, G_0(E) \Delta V \,  G_0 (E)\, \Delta V\, G_0 (E) 
\nonumber \\ 
& = &  \left(\frac{1}{\pi}\right) \mbox{Im}\mbox{Trace}
 \int^{\infty}_{E_{\rm bottom}} dE \,f(E)\, (E - E_F) \, \Delta V \, G_0 (E) \,
\Delta V \, \frac{d}{dE} G_0 (E) \,.
\end{eqnarray}

By performing then the integration by parts, we arrive at

\begin{eqnarray}
{\cal F}^{(2)} & = & 
- \frac{1}{2} \bigg[ \left(\frac{1}{\pi}\right) \mbox{Im}\mbox{Tr} \int^{\infty}_{E_{\rm bottom}} dE\,
                   \frac{d\,f(E)}{dE} \, (E - E_F) \, \Delta V \, G_0 (E) \,
                     \Delta V G_0 (E) \nonumber \\
 &  &+  \left(\frac{1}{\pi}\right) \mbox{Im}\mbox{Tr} \int^{\infty}_{E_{\rm bottom}} dE\,
                   f(E)\, \, \Delta V \, G_0 (E) \,
                     \Delta V\, G_0 (E) \bigg]\,.
\end{eqnarray}

The first term should vanish for the ground state as $\frac{d\,f(E)}{dE}
= \delta (E - E_F)$.

Using the real space multiple scattering representation of the Green
function  $G_0 (\vec{r},\vec{r}',E)$
%eeeeeeeeeeeeeeeeeeeeeeeeeeeeeeeeeeeeeeeeeeeeeeeeeeeeeeeeeeeeeeeeeeeeeeeeeeeeee
  \begin{eqnarray}
G(\vec{r},\vec{r}\,',E) & = &
\sum_{\Lambda_1\Lambda_2} 
Z^{n}_{\Lambda_1}(\vec{r},E)
                              {\tau}^{n n'}_{\Lambda_1\Lambda_2}(E)
Z^{n' \times}_{\Lambda_2}(\vec{r}\,',E) - 
  \sum_{\Lambda_1} \Big[ 
Z^{n}_{\Lambda_1}(\vec{r},E) J^{n \times}_{\Lambda_1}(\vec{r}\,',E)
\Theta(r'-r) \nonumber \\
 & & \;\;\; \;\;  +\;
J^{n}_{\Lambda_1}(\vec{r},E) Z^{n \times}_{\Lambda_1}(\vec{r}\,',E) \Theta(r-r')
\Big] \delta_{nn'} \; 
\label{Eq_KKR-GF}
  \end{eqnarray}
%eeeeeeeeeeeeeeeeeeeeeeeeeeeeeeeeeeeeeeeeeeeeeeeeeeeeeeeeeeeeeeeeeeeeeeeeeeeeee
and the perturbation $\Delta
V$ associated with a change of the spin-dependent potential due to a rotation of
the spin magnetic moment, given by the expression
%eeeeeeeeeeeeeeeeeeeeeeeeeeeeeeeeeeeeeeeeeeeeeeeeeeeeeeeeeeeeeeeeeeeeeeeeeeeeee
\begin{eqnarray}
\Delta V(r) & = & V_{\hat{s}}(r) -  V_{\hat{s}_0}(r) 
  =  \beta (\vec{\sigma} \cdot \delta\hat{s})B(r) \; ,
\label{Eq:Delta_V}
\end{eqnarray}
%eeeeeeeeeeeeeeeeeeeeeeeeeeeeeeeeeeeeeeeeeeeeeeeeeeeeeeeeeeeeeeeeeeeeeeeeeeeeee
(where $\delta\hat{s} = (\delta \hat{s}^x,\delta \hat{s}^y, \delta
\hat{s}^z)$), one obtains the expression for the exchange coupling
parameters as follows
%eeeeeeeeeeeeeeeeeeeeeeeeeeeeeeeeeeeeeeeeeeeeeeeeeeeeeeeeeeeeeeeeeeeeeeeeeeeeee
\begin{eqnarray}
J_{ij}^{\alpha \beta} & = & \frac{\partial^2 {\cal F}^{(2)}}{\partial
                            \hat{s}^\alpha_i \hat{s}^{\beta}_j}
                            \nonumber \\
 & = &- \frac{1}{2\pi} \bigg[ \mbox{Im}\mbox{Tr} \int^{\infty}_{E_{bottom}} dE\,
       \frac{d\,f(E)}{dE} \, (E - E_F) \,
       \langle \Delta V \rangle_i^\alpha \, \tau_{ij} (E) \,
         \langle \Delta V \rangle_j^\beta \, \tau_{ji} (E) \nonumber \\
 &  &+   \mbox{Im}\mbox{Tr} \int^{\infty}_{E_{bottom}} dE\,
                   f(E)\, \,\langle \Delta V \rangle_i^\alpha \, \tau_{ij} (E) \,
                    \langle \Delta V \rangle_j^\beta \, \tau_{ji} (E) \bigg]\,,
\label{Eq:NEW2_Jij}
\end{eqnarray}
%eeeeeeeeeeeeeeeeeeeeeeeeeeeeeeeeeeeeeeeeeeeeeeeeeeeeeeeeeeeeeeeeeeeeeeeeeeeeee
where Green function in Eq.\ (\ref{Eq_KKR-GF}) in terms of the wave functions
$Z^{n}_{\Lambda}(\vec r,E)$ and $J^{n}_{\Lambda}(\vec r,E)$  and the scattering path operator
${\tau}^{nn'}_{ \Lambda  \Lambda'} (E)$ \cite{EBKM16} was used.

\end{widetext}

Using naive arguments, the time dependent behavior of ${\cal E}^{(2)}$
and as a result of the exchange coupling parameters may be calculated
following the scheme used for ground state, but with the Green function $G(E,t)$,
calculated for each time step using potentials obtained within the
TD-DFT, and using occupation function, $f(E,t)$, calculated for each
time step within  TD-DFT calculations.

\section{Spin-lattice coupling parameters  \label{Appendix2}}

Concerning the responsibility of the spin-lattice interactions for
the transverse spin excitations, we focus next on the impact of the
laser pulse on the behavior of the SLC.
The torque on the magnetic moments due to the SLC, $ \vec{\cal T}_i^{SLC} = \hat{s}_i \times
\vec{H}_{SLC}$, responsible for the creation of transverse spin
fluctuations,  may be associated with different types of contributions,
e.g. by the MCA-like and DMI-like SLC \cite{MLPE23}.
Corresponding effective fields originating from the  MCA-like term in the
spin-lattice Hamiltonian are given in terms of the parameters ${\cal
  K}^{\alpha z,\mu}_{i,k}$ as follows
%eeeeeeeeeeeeeeeeeeeeeeeeeeeeeeeeeeeeeeeeeeeeeeeeeeeeeeeeeeeeeeeeeeeeeeeeeeeeee
\begin{eqnarray}  
{{H}^{\alpha}_{\rm{sl-MA},i}} &=& 
 \sum_{k}  { {K}_{i,k}^{z \alpha,\mu}} (u^\mu_k - u^\mu_i) \, ,
\label{Eq_Torque_MCA_spin}
\end{eqnarray}
% eeeeeeeeeeeeeeeeeeeeeeeeeeeeeeeeeeeeeeeeeeeeeeeeeeeeeeeeeeeeeeeeeeeeeeeeeeeeee
while the DMI-like interactions give rise to the effective field 
%eeeeeeeeeeeeeeeeeeeeeeeeeeeeeeeeeeeeeeeeeeeeeeeeeeeeeeeeeeeeeeeeeeeeeeeeeeeeee
\begin{eqnarray}  
{{H}^{\alpha}_{\rm{sl-DMI},i}} &=& \sum_{j,k,\mu} [\hat{s}_j
                                   \times {\cal D}^{\mu}_{ij,k}]_\alpha (u_k^\mu - u_i^\mu) \,.
\label{Eq_Torque_DMI_spin}
\end{eqnarray}
% eeeeeeeeeeeeeeeeeeeeeeeeeeeeeeeeeeeeeeeeeeeeeeeeeeeeeeeeeeeeeeeeeeeeeeeeeeeeee

On the other hand, the SLC-driven force on atom $k$ due to a spin tilting
on site $i$ in the case of MCA-like SLC is given by the expression

%eeeeeeeeeeeeeeeeeeeeeeeeeeeeeeeeeeeeeeeeeeeeeeeeeeeeeeeeeeeeeeeeeeeeeeeeeeeeee
\begin{eqnarray}  
{{\cal F}^{\mu}_{\rm{sl-MA},k}} &=&  \sum_{\alpha}  { {K}_{i,k}^{z
                                    \alpha,\mu}} \delta\hat{s}^\alpha_i
\label{Eq_Torque_MCA_spin}
\end{eqnarray}
% eeeeeeeeeeeeeeeeeeeeeeeeeeeeeeeeeeeeeeeeeeeeeeeeeeeeeeeeeeeeeeeeeeeeeeeeeeeeee
and in the case of DMI-like SLC the force looks as follows
%eeeeeeeeeeeeeeeeeeeeeeeeeeeeeeeeeeeeeeeeeeeeeeeeeeeeeeeeeeeeeeeeeeeeeeeeeeeeee
\begin{eqnarray}  
{{\cal F}^{\mu}_{\rm{sl-DMI},k}} &=& \sum_{j,\alpha} [\hat{s}_j
                                   \times \vec{\cal D}^{\mu}_{ij,k}]_\alpha \delta\hat{s}^\alpha_i.
\label{Eq_Torque_DMI_spin}
\end{eqnarray}
% eeeeeeeeeeeeeeeeeeeeeeeeeeeeeeeeeeeeeeeeeeeeeeeeeeeeeeeeeeeeeeeeeeeeeeeeeeeeee

Note that the expressions for the SLC parameters ${\cal K}^{\alpha z,\mu}_{i,k}$ and
$\vec{\cal D}^{\mu}_{ij,k}$ derived for the ground state
\cite{MPL+22,MLPE23} should be modified taking into account re-population
of the laser-perturbed electronic states, that can be done in line with
the corresponding modifications given in the Appendix for isotropic
exchange coupling parameters.

% Fig. \ref{fig:Fe_SLC_time} represents  DMI-like and MCA-like SLC
% parameters, $D^{\alpha,\nu}_{01}$ and $K^{\alpha,\nu}_{01}$, changing in
% time during laser pulse, with $\vec{R}_{01} = (-0.5,-0.5,-0.5)a_{lat}$.

%\bibliography{/opt/ak/bib/akhelit,spin-latt_2,HL}

\begin{thebibliography}{46}%
\makeatletter
\providecommand \@ifxundefined [1]{%
 \@ifx{#1\undefined}
}%
\providecommand \@ifnum [1]{%
 \ifnum #1\expandafter \@firstoftwo
 \else \expandafter \@secondoftwo
 \fi
}%
\providecommand \@ifx [1]{%
 \ifx #1\expandafter \@firstoftwo
 \else \expandafter \@secondoftwo
 \fi
}%
\providecommand \natexlab [1]{#1}%
\providecommand \enquote  [1]{``#1''}%
\providecommand \bibnamefont  [1]{#1}%
\providecommand \bibfnamefont [1]{#1}%
\providecommand \citenamefont [1]{#1}%
\providecommand \href@noop [0]{\@secondoftwo}%
\providecommand \href [0]{\begingroup \@sanitize@url \@href}%
\providecommand \@href[1]{\@@startlink{#1}\@@href}%
\providecommand \@@href[1]{\endgroup#1\@@endlink}%
\providecommand \@sanitize@url [0]{\catcode `\\12\catcode `\$12\catcode
  `\&12\catcode `\#12\catcode `\^12\catcode `\_12\catcode `\%12\relax}%
\providecommand \@@startlink[1]{}%
\providecommand \@@endlink[0]{}%
\providecommand \url  [0]{\begingroup\@sanitize@url \@url }%
\providecommand \@url [1]{\endgroup\@href {#1}{\urlprefix }}%
\providecommand \urlprefix  [0]{URL }%
\providecommand \Eprint [0]{\href }%
\providecommand \doibase [0]{http://dx.doi.org/}%
\providecommand \selectlanguage [0]{\@gobble}%
\providecommand \bibinfo  [0]{\@secondoftwo}%
\providecommand \bibfield  [0]{\@secondoftwo}%
\providecommand \translation [1]{[#1]}%
\providecommand \BibitemOpen [0]{}%
\providecommand \bibitemStop [0]{}%
\providecommand \bibitemNoStop [0]{.\EOS\space}%
\providecommand \EOS [0]{\spacefactor3000\relax}%
\providecommand \BibitemShut  [1]{\csname bibitem#1\endcsname}%
\let\auto@bib@innerbib\@empty
%</preamble>
\bibitem [{\citenamefont {Beaurepaire}\ \emph {et~al.}(1996)\citenamefont
  {Beaurepaire}, \citenamefont {Merle}, \citenamefont {Daunois},\ and\
  \citenamefont {Bigot}}]{BMDB96}%
  \BibitemOpen
  \bibfield  {author} {\bibinfo {author} {\bibfnamefont {E.}~\bibnamefont
  {Beaurepaire}}, \bibinfo {author} {\bibfnamefont {J.-C.}\ \bibnamefont
  {Merle}}, \bibinfo {author} {\bibfnamefont {A.}~\bibnamefont {Daunois}}, \
  and\ \bibinfo {author} {\bibfnamefont {J.-Y.}\ \bibnamefont {Bigot}},\ }\href
  {\doibase 10.1103/PhysRevLett.76.4250} {\bibfield  {journal} {\bibinfo
  {journal} {Phys. Rev. Lett.}\ }\textbf {\bibinfo {volume} {76}},\ \bibinfo
  {pages} {4250} (\bibinfo {year} {1996})}\BibitemShut {NoStop}%
\bibitem [{\citenamefont {Kirilyuk}\ \emph {et~al.}(2010)\citenamefont
  {Kirilyuk}, \citenamefont {Kimel},\ and\ \citenamefont {Rasing}}]{KKR10}%
  \BibitemOpen
  \bibfield  {author} {\bibinfo {author} {\bibfnamefont {A.}~\bibnamefont
  {Kirilyuk}}, \bibinfo {author} {\bibfnamefont {A.~V.}\ \bibnamefont {Kimel}},
  \ and\ \bibinfo {author} {\bibfnamefont {T.}~\bibnamefont {Rasing}},\ }\href
  {\doibase 10.1103/RevModPhys.82.2731} {\bibfield  {journal} {\bibinfo
  {journal} {Rev. Mod. Phys.}\ }\textbf {\bibinfo {volume} {82}},\ \bibinfo
  {pages} {2731} (\bibinfo {year} {2010})}\BibitemShut {NoStop}%
\bibitem [{\citenamefont {Bigot}\ and\ \citenamefont {Vomir}(2013)}]{BV13}%
  \BibitemOpen
  \bibfield  {author} {\bibinfo {author} {\bibfnamefont {J.-Y.}\ \bibnamefont
  {Bigot}}\ and\ \bibinfo {author} {\bibfnamefont {M.}~\bibnamefont {Vomir}},\
  }\href {\doibase https://doi.org/10.1002/andp.201200199} {\bibfield
  {journal} {\bibinfo  {journal} {Annalen der Physik}\ }\textbf {\bibinfo
  {volume} {525}},\ \bibinfo {pages} {2} (\bibinfo {year} {2013})},\ \Eprint
  {http://arxiv.org/abs/https://onlinelibrary.wiley.com/doi/pdf/10.1002/andp.201200199}
  {https://onlinelibrary.wiley.com/doi/pdf/10.1002/andp.201200199} \BibitemShut
  {NoStop}%
\bibitem [{\citenamefont {Carva}\ \emph {et~al.}(2017)\citenamefont {Carva},
  \citenamefont {Bal\'a\v{z}},\ and\ \citenamefont {Radu}}]{CBR17}%
  \BibitemOpen
  \bibfield  {author} {\bibinfo {author} {\bibfnamefont {K.}~\bibnamefont
  {Carva}}, \bibinfo {author} {\bibfnamefont {P.}~\bibnamefont {Bal\'a\v{z}}},
  \ and\ \bibinfo {author} {\bibfnamefont {I.}~\bibnamefont {Radu}}\ }(\bibinfo
   {publisher} {Elsevier},\ \bibinfo {year} {2017})\ pp.\ \bibinfo {pages}
  {291--463}\BibitemShut {NoStop}%
\bibitem [{\citenamefont {Bigot}\ \emph {et~al.}(2004)\citenamefont {Bigot},
  \citenamefont {Guidoni}, \citenamefont {Beaurepaire},\ and\ \citenamefont
  {Saeta}}]{BGBS04}%
  \BibitemOpen
  \bibfield  {author} {\bibinfo {author} {\bibfnamefont {J.-Y.}\ \bibnamefont
  {Bigot}}, \bibinfo {author} {\bibfnamefont {L.}~\bibnamefont {Guidoni}},
  \bibinfo {author} {\bibfnamefont {E.}~\bibnamefont {Beaurepaire}}, \ and\
  \bibinfo {author} {\bibfnamefont {P.~N.}\ \bibnamefont {Saeta}},\ }\href
  {\doibase 10.1103/PhysRevLett.93.077401} {\bibfield  {journal} {\bibinfo
  {journal} {Phys. Rev. Lett.}\ }\textbf {\bibinfo {volume} {93}},\ \bibinfo
  {pages} {077401} (\bibinfo {year} {2004})}\BibitemShut {NoStop}%
\bibitem [{\citenamefont {F\"ahnle}\ \emph {et~al.}(2015)\citenamefont
  {F\"ahnle}, \citenamefont {Illg}, \citenamefont {Haag},\ and\ \citenamefont
  {Teeny}}]{FIHT15}%
  \BibitemOpen
  \bibfield  {author} {\bibinfo {author} {\bibfnamefont {M.}~\bibnamefont
  {F\"ahnle}}, \bibinfo {author} {\bibfnamefont {C.}~\bibnamefont {Illg}},
  \bibinfo {author} {\bibfnamefont {M.}~\bibnamefont {Haag}}, \ and\ \bibinfo
  {author} {\bibfnamefont {N.}~\bibnamefont {Teeny}},\ }\href {\doibase
  https://doi.org/10.12693/APhysPolA.127.170} {\bibfield  {journal} {\bibinfo
  {journal} {Acta Phys. Polonika A}\ }\textbf {\bibinfo {volume} {127}},\
  \bibinfo {pages} {170} (\bibinfo {year} {2015})}\BibitemShut {NoStop}%
\bibitem [{\citenamefont {Koopmans}\ \emph {et~al.}(2010)\citenamefont
  {Koopmans}, \citenamefont {Malinowski}, \citenamefont {Dalla~Longa},
  \citenamefont {Steiauf}, \citenamefont {Fähnle}, \citenamefont {Roth},
  \citenamefont {Cinchetti},\ and\ \citenamefont {Aeschlimann}}]{KMD+10}%
  \BibitemOpen
  \bibfield  {author} {\bibinfo {author} {\bibfnamefont {B.}~\bibnamefont
  {Koopmans}}, \bibinfo {author} {\bibfnamefont {G.}~\bibnamefont
  {Malinowski}}, \bibinfo {author} {\bibfnamefont {F.}~\bibnamefont
  {Dalla~Longa}}, \bibinfo {author} {\bibfnamefont {D.}~\bibnamefont
  {Steiauf}}, \bibinfo {author} {\bibfnamefont {M.}~\bibnamefont {Fähnle}},
  \bibinfo {author} {\bibfnamefont {T.}~\bibnamefont {Roth}}, \bibinfo {author}
  {\bibfnamefont {M.}~\bibnamefont {Cinchetti}}, \ and\ \bibinfo {author}
  {\bibfnamefont {M.}~\bibnamefont {Aeschlimann}},\ }\href
  {https://doi.org/10.1038/nmat2593} {\bibfield  {journal} {\bibinfo  {journal}
  {Nature Materials}\ }\textbf {\bibinfo {volume} {9}},\ \bibinfo {pages} {259}
  (\bibinfo {year} {2010})}\BibitemShut {NoStop}%
\bibitem [{\citenamefont {H\"ubner}\ and\ \citenamefont
  {Bennemann}(1996)}]{HB96}%
  \BibitemOpen
  \bibfield  {author} {\bibinfo {author} {\bibfnamefont {W.}~\bibnamefont
  {H\"ubner}}\ and\ \bibinfo {author} {\bibfnamefont {K.~H.}\ \bibnamefont
  {Bennemann}},\ }\href {\doibase 10.1103/PhysRevB.53.3422} {\bibfield
  {journal} {\bibinfo  {journal} {Phys. Rev. B}\ }\textbf {\bibinfo {volume}
  {53}},\ \bibinfo {pages} {3422} (\bibinfo {year} {1996})}\BibitemShut
  {NoStop}%
\bibitem [{\citenamefont {Dornes}\ \emph {et~al.}(2019)\citenamefont {Dornes},
  \citenamefont {Acremann}, \citenamefont {Savoini}, \citenamefont {Kubli},
  \citenamefont {Neugebauer}, \citenamefont {Abreu}, \citenamefont {Huber},
  \citenamefont {Lantz}, \citenamefont {Vaz}, \citenamefont {Lemke},
  \citenamefont {Bothschafter}, \citenamefont {Porer}, \citenamefont
  {Esposito}, \citenamefont {Rettig}, \citenamefont {Buzzi}, \citenamefont
  {Alberca}, \citenamefont {Windsor}, \citenamefont {Beaud}, \citenamefont
  {Staub}, \citenamefont {Zhu}, \citenamefont {Song}, \citenamefont {Glownia},\
  and\ \citenamefont {Johnson}}]{DAS+19}%
  \BibitemOpen
  \bibfield  {author} {\bibinfo {author} {\bibfnamefont {C.}~\bibnamefont
  {Dornes}}, \bibinfo {author} {\bibfnamefont {Y.}~\bibnamefont {Acremann}},
  \bibinfo {author} {\bibfnamefont {M.}~\bibnamefont {Savoini}}, \bibinfo
  {author} {\bibfnamefont {M.}~\bibnamefont {Kubli}}, \bibinfo {author}
  {\bibfnamefont {M.~J.}\ \bibnamefont {Neugebauer}}, \bibinfo {author}
  {\bibfnamefont {E.}~\bibnamefont {Abreu}}, \bibinfo {author} {\bibfnamefont
  {L.}~\bibnamefont {Huber}}, \bibinfo {author} {\bibfnamefont
  {G.}~\bibnamefont {Lantz}}, \bibinfo {author} {\bibfnamefont {C.~A.~F.}\
  \bibnamefont {Vaz}}, \bibinfo {author} {\bibfnamefont {H.}~\bibnamefont
  {Lemke}}, \bibinfo {author} {\bibfnamefont {E.~M.}\ \bibnamefont
  {Bothschafter}}, \bibinfo {author} {\bibfnamefont {M.}~\bibnamefont {Porer}},
  \bibinfo {author} {\bibfnamefont {V.}~\bibnamefont {Esposito}}, \bibinfo
  {author} {\bibfnamefont {L.}~\bibnamefont {Rettig}}, \bibinfo {author}
  {\bibfnamefont {M.}~\bibnamefont {Buzzi}}, \bibinfo {author} {\bibfnamefont
  {A.}~\bibnamefont {Alberca}}, \bibinfo {author} {\bibfnamefont {Y.~W.}\
  \bibnamefont {Windsor}}, \bibinfo {author} {\bibfnamefont {P.}~\bibnamefont
  {Beaud}}, \bibinfo {author} {\bibfnamefont {U.}~\bibnamefont {Staub}},
  \bibinfo {author} {\bibfnamefont {D.}~\bibnamefont {Zhu}}, \bibinfo {author}
  {\bibfnamefont {S.}~\bibnamefont {Song}}, \bibinfo {author} {\bibfnamefont
  {J.~M.}\ \bibnamefont {Glownia}}, \ and\ \bibinfo {author} {\bibfnamefont
  {S.~L.}\ \bibnamefont {Johnson}},\ }\href
  {https://doi.org/10.1038/s41586-018-0822-7} {\bibfield  {journal} {\bibinfo
  {journal} {Nature}\ }\textbf {\bibinfo {volume} {565}},\ \bibinfo {pages}
  {209} (\bibinfo {year} {2019})}\BibitemShut {NoStop}%
\bibitem [{\citenamefont {Zhang}\ and\ \citenamefont {George}(2008)}]{ZG08}%
  \BibitemOpen
  \bibfield  {author} {\bibinfo {author} {\bibfnamefont {G.~P.}\ \bibnamefont
  {Zhang}}\ and\ \bibinfo {author} {\bibfnamefont {T.~F.}\ \bibnamefont
  {George}},\ }\href {\doibase 10.1103/PhysRevB.78.052407} {\bibfield
  {journal} {\bibinfo  {journal} {Phys. Rev. B}\ }\textbf {\bibinfo {volume}
  {78}},\ \bibinfo {pages} {052407} (\bibinfo {year} {2008})}\BibitemShut
  {NoStop}%
\bibitem [{\citenamefont {Zhang}\ and\ \citenamefont {H\"ubner}(2000)}]{ZH00}%
  \BibitemOpen
  \bibfield  {author} {\bibinfo {author} {\bibfnamefont {G.~P.}\ \bibnamefont
  {Zhang}}\ and\ \bibinfo {author} {\bibfnamefont {W.}~\bibnamefont
  {H\"ubner}},\ }\href {\doibase 10.1103/PhysRevLett.85.3025} {\bibfield
  {journal} {\bibinfo  {journal} {Phys. Rev. Lett.}\ }\textbf {\bibinfo
  {volume} {85}},\ \bibinfo {pages} {3025} (\bibinfo {year}
  {2000})}\BibitemShut {NoStop}%
\bibitem [{\citenamefont {Carva}\ \emph {et~al.}(2011)\citenamefont {Carva},
  \citenamefont {Battiato},\ and\ \citenamefont {Oppeneer}}]{CBO11}%
  \BibitemOpen
  \bibfield  {author} {\bibinfo {author} {\bibfnamefont {K.}~\bibnamefont
  {Carva}}, \bibinfo {author} {\bibfnamefont {M.}~\bibnamefont {Battiato}}, \
  and\ \bibinfo {author} {\bibfnamefont {P.~M.}\ \bibnamefont {Oppeneer}},\
  }\href {\doibase 10.1103/PhysRevLett.107.207201} {\bibfield  {journal}
  {\bibinfo  {journal} {Phys. Rev. Lett.}\ }\textbf {\bibinfo {volume} {107}},\
  \bibinfo {pages} {207201} (\bibinfo {year} {2011})}\BibitemShut {NoStop}%
\bibitem [{\citenamefont {Illg}\ \emph {et~al.}(2013)\citenamefont {Illg},
  \citenamefont {Haag},\ and\ \citenamefont {F\"ahnle}}]{IHF13}%
  \BibitemOpen
  \bibfield  {author} {\bibinfo {author} {\bibfnamefont {C.}~\bibnamefont
  {Illg}}, \bibinfo {author} {\bibfnamefont {M.}~\bibnamefont {Haag}}, \ and\
  \bibinfo {author} {\bibfnamefont {M.}~\bibnamefont {F\"ahnle}},\ }\href
  {\doibase 10.1103/PhysRevB.88.214404} {\bibfield  {journal} {\bibinfo
  {journal} {Phys. Rev. B}\ }\textbf {\bibinfo {volume} {88}},\ \bibinfo
  {pages} {214404} (\bibinfo {year} {2013})}\BibitemShut {NoStop}%
\bibitem [{\citenamefont {Carva}\ \emph {et~al.}(2015)\citenamefont {Carva},
  \citenamefont {Battiato}, \citenamefont {Legut},\ and\ \citenamefont
  {Oppeneer}}]{CBLO15}%
  \BibitemOpen
  \bibfield  {author} {\bibinfo {author} {\bibfnamefont {K.}~\bibnamefont
  {Carva}}, \bibinfo {author} {\bibfnamefont {M.}~\bibnamefont {Battiato}},
  \bibinfo {author} {\bibfnamefont {D.}~\bibnamefont {Legut}}, \ and\ \bibinfo
  {author} {\bibfnamefont {P.~M.}\ \bibnamefont {Oppeneer}},\ }\enquote
  {\bibinfo {title} {Theory of femtosecond laser-induced demagnetization},}\
  in\ \href@noop {} {\emph {\bibinfo {booktitle} {Ultrafast Magnetism I:
  Proceedings of the International Conference UMC 2013 Strasbourg, France}}}\
  (\bibinfo  {publisher} {Springer International Publishing Switzerland 2015},\
  \bibinfo {address} {Cham, Heidelberg, New York, Dordrecht, London},\ \bibinfo
  {year} {2015})\ pp.\ \bibinfo {pages} {111--115}\BibitemShut {NoStop}%
\bibitem [{\citenamefont {{Krieger}}\ \emph {et~al.}(2014)\citenamefont
  {{Krieger}}, \citenamefont {{Dewhurst}}, \citenamefont {{Elliott}},
  \citenamefont {{Sharma}},\ and\ \citenamefont {{Gross}}}]{KDE+14}%
  \BibitemOpen
  \bibfield  {author} {\bibinfo {author} {\bibfnamefont {K.}~\bibnamefont
  {{Krieger}}}, \bibinfo {author} {\bibfnamefont {J.~K.}\ \bibnamefont
  {{Dewhurst}}}, \bibinfo {author} {\bibfnamefont {P.}~\bibnamefont
  {{Elliott}}}, \bibinfo {author} {\bibfnamefont {S.}~\bibnamefont {{Sharma}}},
  \ and\ \bibinfo {author} {\bibfnamefont {E.~K.~U.}\ \bibnamefont {{Gross}}},\
  }\href@noop {} {\bibfield  {journal} {\bibinfo  {journal} {ArXiv e-prints,
  1406.6607}\ } (\bibinfo {year} {2014})},\ \Eprint
  {http://arxiv.org/abs/1406.6607} {arXiv:1406.6607 [cond-mat.mtrl-sci]}
  \BibitemShut {NoStop}%
\bibitem [{\citenamefont {Sharma}\ \emph {et~al.}(2022)\citenamefont {Sharma},
  \citenamefont {Shallcross}, \citenamefont {Elliott},\ and\ \citenamefont
  {Dewhurst}}]{SSED22}%
  \BibitemOpen
  \bibfield  {author} {\bibinfo {author} {\bibfnamefont {S.}~\bibnamefont
  {Sharma}}, \bibinfo {author} {\bibfnamefont {S.}~\bibnamefont {Shallcross}},
  \bibinfo {author} {\bibfnamefont {P.}~\bibnamefont {Elliott}}, \ and\
  \bibinfo {author} {\bibfnamefont {J.~K.}\ \bibnamefont {Dewhurst}},\
  }\href@noop {} {\bibfield  {journal} {\bibinfo  {journal} {Science advances}\
  }\textbf {\bibinfo {volume} {8}},\ \bibinfo {pages} {eabq2021} (\bibinfo
  {year} {2022})}\BibitemShut {NoStop}%
\bibitem [{\citenamefont {Schellekens}\ and\ \citenamefont
  {Koopmans}(2013)}]{SK13}%
  \BibitemOpen
  \bibfield  {author} {\bibinfo {author} {\bibfnamefont {A.~J.}\ \bibnamefont
  {Schellekens}}\ and\ \bibinfo {author} {\bibfnamefont {B.}~\bibnamefont
  {Koopmans}},\ }\href {\doibase 10.1103/PhysRevLett.110.217204} {\bibfield
  {journal} {\bibinfo  {journal} {Phys. Rev. Lett.}\ }\textbf {\bibinfo
  {volume} {110}},\ \bibinfo {pages} {217204} (\bibinfo {year}
  {2013})}\BibitemShut {NoStop}%
\bibitem [{\citenamefont {F\"ahnle}\ \emph {et~al.}(2017)\citenamefont
  {F\"ahnle}, \citenamefont {Tsatsoulis}, \citenamefont {Illg}, \citenamefont
  {Haag}, \citenamefont {M\"uller},\ and\ \citenamefont {Zhang}}]{FTI+17}%
  \BibitemOpen
  \bibfield  {author} {\bibinfo {author} {\bibfnamefont {M.}~\bibnamefont
  {F\"ahnle}}, \bibinfo {author} {\bibfnamefont {T.}~\bibnamefont
  {Tsatsoulis}}, \bibinfo {author} {\bibfnamefont {C.}~\bibnamefont {Illg}},
  \bibinfo {author} {\bibfnamefont {M.}~\bibnamefont {Haag}}, \bibinfo {author}
  {\bibfnamefont {B.~Y.}\ \bibnamefont {M\"uller}}, \ and\ \bibinfo {author}
  {\bibfnamefont {L.}~\bibnamefont {Zhang}},\ }\href
  {https://doi.org/10.1007/s10948-016-3950-z} {\bibfield  {journal} {\bibinfo
  {journal} {Journal of Superconductivity and Novel Magnetism}\ }\textbf
  {\bibinfo {volume} {30}},\ \bibinfo {pages} {1381} (\bibinfo {year}
  {2017})}\BibitemShut {NoStop}%
\bibitem [{\citenamefont {Eich}\ \emph {et~al.}(2017)\citenamefont {Eich},
  \citenamefont {Pl\"otzing}, \citenamefont {Rollinger}, \citenamefont
  {Emmerich}, \citenamefont {Adam}, \citenamefont {Chen}, \citenamefont
  {Kapteyn}, \citenamefont {Murnane}, \citenamefont {Plucinski}, \citenamefont
  {Steil}, \citenamefont {Stadtm\"uller}, \citenamefont {Cinchetti},
  \citenamefont {Aeschlimann}, \citenamefont {Schneider},\ and\ \citenamefont
  {Mathias}}]{EPR+17}%
  \BibitemOpen
  \bibfield  {author} {\bibinfo {author} {\bibfnamefont {S.}~\bibnamefont
  {Eich}}, \bibinfo {author} {\bibfnamefont {M.}~\bibnamefont {Pl\"otzing}},
  \bibinfo {author} {\bibfnamefont {M.}~\bibnamefont {Rollinger}}, \bibinfo
  {author} {\bibfnamefont {S.}~\bibnamefont {Emmerich}}, \bibinfo {author}
  {\bibfnamefont {R.}~\bibnamefont {Adam}}, \bibinfo {author} {\bibfnamefont
  {C.}~\bibnamefont {Chen}}, \bibinfo {author} {\bibfnamefont {H.~C.}\
  \bibnamefont {Kapteyn}}, \bibinfo {author} {\bibfnamefont {M.~M.}\
  \bibnamefont {Murnane}}, \bibinfo {author} {\bibfnamefont {L.}~\bibnamefont
  {Plucinski}}, \bibinfo {author} {\bibfnamefont {D.}~\bibnamefont {Steil}},
  \bibinfo {author} {\bibfnamefont {B.}~\bibnamefont {Stadtm\"uller}}, \bibinfo
  {author} {\bibfnamefont {M.}~\bibnamefont {Cinchetti}}, \bibinfo {author}
  {\bibfnamefont {M.}~\bibnamefont {Aeschlimann}}, \bibinfo {author}
  {\bibfnamefont {C.~M.}\ \bibnamefont {Schneider}}, \ and\ \bibinfo {author}
  {\bibfnamefont {S.}~\bibnamefont {Mathias}},\ }\href {\doibase
  10.1126/sciadv.1602094} {\bibfield  {journal} {\bibinfo  {journal} {Science
  Advances}\ }\textbf {\bibinfo {volume} {3}} (\bibinfo {year} {2017}),\
  10.1126/sciadv.1602094},\ \Eprint
  {http://arxiv.org/abs/http://advances.sciencemag.org/content/3/3/e1602094.full.pdf}
  {http://advances.sciencemag.org/content/3/3/e1602094.full.pdf} \BibitemShut
  {NoStop}%
\bibitem [{\citenamefont {Tengdin}\ \emph {et~al.}(2018)\citenamefont
  {Tengdin}, \citenamefont {You}, \citenamefont {Chen}, \citenamefont {Shi},
  \citenamefont {Zusin}, \citenamefont {Zhang}, \citenamefont {Gentry},
  \citenamefont {Blonsky}, \citenamefont {Keller}, \citenamefont {Oppeneer},
  \citenamefont {Kapteyn}, \citenamefont {Tao},\ and\ \citenamefont
  {Murnane}}]{TYC+18}%
  \BibitemOpen
  \bibfield  {author} {\bibinfo {author} {\bibfnamefont {P.}~\bibnamefont
  {Tengdin}}, \bibinfo {author} {\bibfnamefont {W.}~\bibnamefont {You}},
  \bibinfo {author} {\bibfnamefont {C.}~\bibnamefont {Chen}}, \bibinfo {author}
  {\bibfnamefont {X.}~\bibnamefont {Shi}}, \bibinfo {author} {\bibfnamefont
  {D.}~\bibnamefont {Zusin}}, \bibinfo {author} {\bibfnamefont
  {Y.}~\bibnamefont {Zhang}}, \bibinfo {author} {\bibfnamefont
  {C.}~\bibnamefont {Gentry}}, \bibinfo {author} {\bibfnamefont
  {A.}~\bibnamefont {Blonsky}}, \bibinfo {author} {\bibfnamefont
  {M.}~\bibnamefont {Keller}}, \bibinfo {author} {\bibfnamefont {P.~M.}\
  \bibnamefont {Oppeneer}}, \bibinfo {author} {\bibfnamefont {H.~C.}\
  \bibnamefont {Kapteyn}}, \bibinfo {author} {\bibfnamefont {Z.}~\bibnamefont
  {Tao}}, \ and\ \bibinfo {author} {\bibfnamefont {M.~M.}\ \bibnamefont
  {Murnane}},\ }\href {\doibase 10.1126/sciadv.aap9744} {\bibfield  {journal}
  {\bibinfo  {journal} {Science Advances}\ }\textbf {\bibinfo {volume} {4}},\
  \bibinfo {pages} {eaap9744} (\bibinfo {year} {2018})},\ \Eprint
  {http://arxiv.org/abs/https://www.science.org/doi/pdf/10.1126/sciadv.aap9744}
  {https://www.science.org/doi/pdf/10.1126/sciadv.aap9744} \BibitemShut
  {NoStop}%
\bibitem [{\citenamefont {Kazantseva}\ \emph {et~al.}(2007)\citenamefont
  {Kazantseva}, \citenamefont {Nowak}, \citenamefont {Chantrell}, \citenamefont
  {Hohlfeld},\ and\ \citenamefont {Rebei}}]{KNC+08}%
  \BibitemOpen
  \bibfield  {author} {\bibinfo {author} {\bibfnamefont {N.}~\bibnamefont
  {Kazantseva}}, \bibinfo {author} {\bibfnamefont {U.}~\bibnamefont {Nowak}},
  \bibinfo {author} {\bibfnamefont {R.~W.}\ \bibnamefont {Chantrell}}, \bibinfo
  {author} {\bibfnamefont {J.}~\bibnamefont {Hohlfeld}}, \ and\ \bibinfo
  {author} {\bibfnamefont {A.}~\bibnamefont {Rebei}},\ }\href {\doibase
  10.1209/0295-5075/81/27004} {\bibfield  {journal} {\bibinfo  {journal}
  {Europhysics Letters}\ }\textbf {\bibinfo {volume} {81}},\ \bibinfo {pages}
  {27004} (\bibinfo {year} {2007})}\BibitemShut {NoStop}%
\bibitem [{\citenamefont {De\'ak}\ \emph {et~al.}(2017)\citenamefont {De\'ak},
  \citenamefont {Hinzke}, \citenamefont {Szunyogh},\ and\ \citenamefont
  {Nowak}}]{DHSN17}%
  \BibitemOpen
  \bibfield  {author} {\bibinfo {author} {\bibfnamefont {A.}~\bibnamefont
  {De\'ak}}, \bibinfo {author} {\bibfnamefont {D.}~\bibnamefont {Hinzke}},
  \bibinfo {author} {\bibfnamefont {L.}~\bibnamefont {Szunyogh}}, \ and\
  \bibinfo {author} {\bibfnamefont {U.}~\bibnamefont {Nowak}},\ }\href
  {\doibase 10.1088/1361-648X/aa76fc} {\bibfield  {journal} {\bibinfo
  {journal} {Journal of Physics: Condensed Matter}\ }\textbf {\bibinfo {volume}
  {29}},\ \bibinfo {pages} {314003} (\bibinfo {year} {2017})}\BibitemShut
  {NoStop}%
\bibitem [{\citenamefont {Weißenhofer}\ and\ \citenamefont
  {Oppeneer}(2023)}]{WO23}%
  \BibitemOpen
  \bibfield  {author} {\bibinfo {author} {\bibfnamefont {M.}~\bibnamefont
  {Weißenhofer}}\ and\ \bibinfo {author} {\bibfnamefont {M.~P.}\ \bibnamefont
  {Oppeneer}},\ }\href {\doibase 10.1002/apxr.202300103} {\bibfield  {journal}
  {\bibinfo  {journal} {arXiv}\ ,\ \bibinfo {pages} {2309.14167v3}} (\bibinfo
  {year} {2023})}\BibitemShut {NoStop}%
\bibitem [{\citenamefont {Němec}\ \emph {et~al.}(2012)\citenamefont {Němec},
  \citenamefont {Rozkotová}, \citenamefont {Tesařová}, \citenamefont
  {Trojánek}, \citenamefont {De~Ranieri}, \citenamefont {Olejník},
  \citenamefont {Zemen}, \citenamefont {Novák}, \citenamefont {Cukr},
  \citenamefont {Malý},\ and\ \citenamefont {Jungwirth}}]{NRT+12}%
  \BibitemOpen
  \bibfield  {author} {\bibinfo {author} {\bibfnamefont {P.}~\bibnamefont
  {Němec}}, \bibinfo {author} {\bibfnamefont {E.}~\bibnamefont {Rozkotová}},
  \bibinfo {author} {\bibfnamefont {N.}~\bibnamefont {Tesařová}}, \bibinfo
  {author} {\bibfnamefont {F.}~\bibnamefont {Trojánek}}, \bibinfo {author}
  {\bibfnamefont {E.}~\bibnamefont {De~Ranieri}}, \bibinfo {author}
  {\bibfnamefont {K.}~\bibnamefont {Olejník}}, \bibinfo {author}
  {\bibfnamefont {J.}~\bibnamefont {Zemen}}, \bibinfo {author} {\bibfnamefont
  {V.}~\bibnamefont {Novák}}, \bibinfo {author} {\bibfnamefont
  {M.}~\bibnamefont {Cukr}}, \bibinfo {author} {\bibfnamefont {P.}~\bibnamefont
  {Malý}}, \ and\ \bibinfo {author} {\bibfnamefont {T.}~\bibnamefont
  {Jungwirth}},\ }\href {https://doi.org/10.1038/nphys2279} {\bibfield
  {journal} {\bibinfo  {journal} {Nature Physics}\ }\textbf {\bibinfo {volume}
  {8}},\ \bibinfo {pages} {411} (\bibinfo {year} {2012})}\BibitemShut {NoStop}%
\bibitem [{\citenamefont {Subkhangulov}\ \emph {et~al.}(2014)\citenamefont
  {Subkhangulov}, \citenamefont {Henriques}, \citenamefont {Rappl},
  \citenamefont {Abramof}, \citenamefont {Rasing},\ and\ \citenamefont
  {Kimel}}]{SHR+14}%
  \BibitemOpen
  \bibfield  {author} {\bibinfo {author} {\bibfnamefont {R.~R.}\ \bibnamefont
  {Subkhangulov}}, \bibinfo {author} {\bibfnamefont {A.~B.}\ \bibnamefont
  {Henriques}}, \bibinfo {author} {\bibfnamefont {P.~H.~O.}\ \bibnamefont
  {Rappl}}, \bibinfo {author} {\bibfnamefont {E.}~\bibnamefont {Abramof}},
  \bibinfo {author} {\bibfnamefont {T.}~\bibnamefont {Rasing}}, \ and\ \bibinfo
  {author} {\bibfnamefont {A.~V.}\ \bibnamefont {Kimel}},\ }\href
  {https://doi.org/10.1038/srep04368} {\bibfield  {journal} {\bibinfo
  {journal} {Scientific Reports}\ }\textbf {\bibinfo {volume} {4}},\ \bibinfo
  {pages} {4368} (\bibinfo {year} {2014})}\BibitemShut {NoStop}%
\bibitem [{\citenamefont {Mikhaylovskiy}\ \emph {et~al.}(2015)\citenamefont
  {Mikhaylovskiy}, \citenamefont {Hendry}, \citenamefont {Secchi},
  \citenamefont {Mentink}, \citenamefont {Eckstein}, \citenamefont {Wu},
  \citenamefont {Pisarev}, \citenamefont {Kruglyak}, \citenamefont
  {Katsnelson}, \citenamefont {Rasing},\ and\ \citenamefont {Kimel}}]{MHS+15}%
  \BibitemOpen
  \bibfield  {author} {\bibinfo {author} {\bibfnamefont {R.~V.}\ \bibnamefont
  {Mikhaylovskiy}}, \bibinfo {author} {\bibfnamefont {E.}~\bibnamefont
  {Hendry}}, \bibinfo {author} {\bibfnamefont {A.}~\bibnamefont {Secchi}},
  \bibinfo {author} {\bibfnamefont {J.~H.}\ \bibnamefont {Mentink}}, \bibinfo
  {author} {\bibfnamefont {M.}~\bibnamefont {Eckstein}}, \bibinfo {author}
  {\bibfnamefont {A.}~\bibnamefont {Wu}}, \bibinfo {author} {\bibfnamefont
  {R.~V.}\ \bibnamefont {Pisarev}}, \bibinfo {author} {\bibfnamefont {V.~V.}\
  \bibnamefont {Kruglyak}}, \bibinfo {author} {\bibfnamefont {M.~I.}\
  \bibnamefont {Katsnelson}}, \bibinfo {author} {\bibfnamefont
  {T.}~\bibnamefont {Rasing}}, \ and\ \bibinfo {author} {\bibfnamefont {A.~V.}\
  \bibnamefont {Kimel}},\ }\href {https://doi.org/10.1038/ncomms9190}
  {\bibfield  {journal} {\bibinfo  {journal} {Nature Communications}\ }\textbf
  {\bibinfo {volume} {6}},\ \bibinfo {pages} {8190} (\bibinfo {year}
  {2015})}\BibitemShut {NoStop}%
\bibitem [{\citenamefont {Freimuth}\ \emph {et~al.}(2016)\citenamefont
  {Freimuth}, \citenamefont {Bl\"ugel},\ and\ \citenamefont
  {Mokrousov}}]{FBM16}%
  \BibitemOpen
  \bibfield  {author} {\bibinfo {author} {\bibfnamefont {F.}~\bibnamefont
  {Freimuth}}, \bibinfo {author} {\bibfnamefont {S.}~\bibnamefont {Bl\"ugel}},
  \ and\ \bibinfo {author} {\bibfnamefont {Y.}~\bibnamefont {Mokrousov}},\
  }\href {\doibase 10.1103/PhysRevB.94.144432} {\bibfield  {journal} {\bibinfo
  {journal} {Phys. Rev. B}\ }\textbf {\bibinfo {volume} {94}},\ \bibinfo
  {pages} {144432} (\bibinfo {year} {2016})}\BibitemShut {NoStop}%
\bibitem [{\citenamefont {Mentink}(2017)}]{Men17}%
  \BibitemOpen
  \bibfield  {author} {\bibinfo {author} {\bibfnamefont {J.~H.}\ \bibnamefont
  {Mentink}},\ }\href {\doibase 10.1088/1361-648X/aa8abf} {\bibfield  {journal}
  {\bibinfo  {journal} {Journal of Physics: Condensed Matter}\ }\textbf
  {\bibinfo {volume} {29}},\ \bibinfo {pages} {453001} (\bibinfo {year}
  {2017})}\BibitemShut {NoStop}%
\bibitem [{\citenamefont {Hamamera}\ \emph {et~al.}(2022)\citenamefont
  {Hamamera}, \citenamefont {Guimarães}, \citenamefont {dos Santos~Dias},\
  and\ \citenamefont {Lounis}}]{HGSL22}%
  \BibitemOpen
  \bibfield  {author} {\bibinfo {author} {\bibfnamefont {H.}~\bibnamefont
  {Hamamera}}, \bibinfo {author} {\bibfnamefont {F.~S.~M.}\ \bibnamefont
  {Guimarães}}, \bibinfo {author} {\bibfnamefont {M.}~\bibnamefont {dos
  Santos~Dias}}, \ and\ \bibinfo {author} {\bibfnamefont {S.}~\bibnamefont
  {Lounis}},\ }\href {https://doi.org/10.1038/s42005-021-00798-8} {\bibfield
  {journal} {\bibinfo  {journal} {Communications Physics}\ }\textbf {\bibinfo
  {volume} {5}},\ \bibinfo {pages} {16} (\bibinfo {year} {2022})}\BibitemShut
  {NoStop}%
\bibitem [{\citenamefont {Núñez}\ \emph {et~al.}(2004)\citenamefont
  {Núñez}, \citenamefont {Fernández-Rossier}, \citenamefont {Abolfath},\
  and\ \citenamefont {MacDonald}}]{NFAM04}%
  \BibitemOpen
  \bibfield  {author} {\bibinfo {author} {\bibfnamefont {A.~S.}\ \bibnamefont
  {Núñez}}, \bibinfo {author} {\bibfnamefont {J.}~\bibnamefont
  {Fernández-Rossier}}, \bibinfo {author} {\bibfnamefont {M.}~\bibnamefont
  {Abolfath}}, \ and\ \bibinfo {author} {\bibfnamefont {A.}~\bibnamefont
  {MacDonald}},\ }\href {\doibase https://doi.org/10.1016/j.jmmm.2003.12.1168}
  {\bibfield  {journal} {\bibinfo  {journal} {Journal of Magnetism and Magnetic
  Materials}\ }\textbf {\bibinfo {volume} {272-276}},\ \bibinfo {pages} {1913}
  (\bibinfo {year} {2004})},\ \bibinfo {note} {proceedings of the International
  Conference on Magnetism (ICM 2003)}\BibitemShut {NoStop}%
\bibitem [{\citenamefont {R\"uckriegel}\ \emph {et~al.}(2020)\citenamefont
  {R\"uckriegel}, \citenamefont {Streib}, \citenamefont {Bauer},\ and\
  \citenamefont {Duine}}]{RSBD20}%
  \BibitemOpen
  \bibfield  {author} {\bibinfo {author} {\bibfnamefont {A.}~\bibnamefont
  {R\"uckriegel}}, \bibinfo {author} {\bibfnamefont {S.}~\bibnamefont
  {Streib}}, \bibinfo {author} {\bibfnamefont {G.~E.~W.}\ \bibnamefont
  {Bauer}}, \ and\ \bibinfo {author} {\bibfnamefont {R.~A.}\ \bibnamefont
  {Duine}},\ }\href {\doibase 10.1103/PhysRevB.101.104402} {\bibfield
  {journal} {\bibinfo  {journal} {Phys. Rev. B}\ }\textbf {\bibinfo {volume}
  {101}},\ \bibinfo {pages} {104402} (\bibinfo {year} {2020})}\BibitemShut
  {NoStop}%
\bibitem [{\citenamefont {Frietsch}\ \emph {et~al.}(2020)\citenamefont
  {Frietsch}, \citenamefont {Donges}, \citenamefont {Carley}, \citenamefont
  {Teichmann}, \citenamefont {Bowlan}, \citenamefont {Döbrich}, \citenamefont
  {Carva}, \citenamefont {Legut}, \citenamefont {Oppeneer}, \citenamefont
  {Nowak},\ and\ \citenamefont {Weinelt}}]{FDC+20}%
  \BibitemOpen
  \bibfield  {author} {\bibinfo {author} {\bibfnamefont {B.}~\bibnamefont
  {Frietsch}}, \bibinfo {author} {\bibfnamefont {A.}~\bibnamefont {Donges}},
  \bibinfo {author} {\bibfnamefont {R.}~\bibnamefont {Carley}}, \bibinfo
  {author} {\bibfnamefont {M.}~\bibnamefont {Teichmann}}, \bibinfo {author}
  {\bibfnamefont {J.}~\bibnamefont {Bowlan}}, \bibinfo {author} {\bibfnamefont
  {K.}~\bibnamefont {Döbrich}}, \bibinfo {author} {\bibfnamefont
  {K.}~\bibnamefont {Carva}}, \bibinfo {author} {\bibfnamefont
  {D.}~\bibnamefont {Legut}}, \bibinfo {author} {\bibfnamefont {P.~M.}\
  \bibnamefont {Oppeneer}}, \bibinfo {author} {\bibfnamefont {U.}~\bibnamefont
  {Nowak}}, \ and\ \bibinfo {author} {\bibfnamefont {M.}~\bibnamefont
  {Weinelt}},\ }\href {\doibase 10.1126/sciadv.abb1601} {\bibfield  {journal}
  {\bibinfo  {journal} {Science Advances}\ }\textbf {\bibinfo {volume} {6}},\
  \bibinfo {pages} {eabb1601} (\bibinfo {year} {2020})},\ \Eprint
  {http://arxiv.org/abs/https://www.science.org/doi/pdf/10.1126/sciadv.abb1601}
  {https://www.science.org/doi/pdf/10.1126/sciadv.abb1601} \BibitemShut
  {NoStop}%
\bibitem [{\citenamefont {van Kampen}\ \emph {et~al.}(2002)\citenamefont {van
  Kampen}, \citenamefont {Jozsa}, \citenamefont {Kohlhepp}, \citenamefont
  {LeClair}, \citenamefont {Lagae}, \citenamefont {de~Jonge},\ and\
  \citenamefont {Koopmans}}]{KJK+02}%
  \BibitemOpen
  \bibfield  {author} {\bibinfo {author} {\bibfnamefont {M.}~\bibnamefont {van
  Kampen}}, \bibinfo {author} {\bibfnamefont {C.}~\bibnamefont {Jozsa}},
  \bibinfo {author} {\bibfnamefont {J.~T.}\ \bibnamefont {Kohlhepp}}, \bibinfo
  {author} {\bibfnamefont {P.}~\bibnamefont {LeClair}}, \bibinfo {author}
  {\bibfnamefont {L.}~\bibnamefont {Lagae}}, \bibinfo {author} {\bibfnamefont
  {W.~J.~M.}\ \bibnamefont {de~Jonge}}, \ and\ \bibinfo {author} {\bibfnamefont
  {B.}~\bibnamefont {Koopmans}},\ }\href {\doibase
  10.1103/PhysRevLett.88.227201} {\bibfield  {journal} {\bibinfo  {journal}
  {Phys. Rev. Lett.}\ }\textbf {\bibinfo {volume} {88}},\ \bibinfo {pages}
  {227201} (\bibinfo {year} {2002})}\BibitemShut {NoStop}%
\bibitem [{\citenamefont {Secchi}\ \emph {et~al.}(2013)\citenamefont {Secchi},
  \citenamefont {Brener}, \citenamefont {Lichtenstein},\ and\ \citenamefont
  {Katsnelson}}]{SBLK13}%
  \BibitemOpen
  \bibfield  {author} {\bibinfo {author} {\bibfnamefont {A.}~\bibnamefont
  {Secchi}}, \bibinfo {author} {\bibfnamefont {S.}~\bibnamefont {Brener}},
  \bibinfo {author} {\bibfnamefont {A.}~\bibnamefont {Lichtenstein}}, \ and\
  \bibinfo {author} {\bibfnamefont {M.}~\bibnamefont {Katsnelson}},\ }\href
  {\doibase https://doi.org/10.1016/j.aop.2013.03.006} {\bibfield  {journal}
  {\bibinfo  {journal} {Annals of Physics}\ }\textbf {\bibinfo {volume}
  {333}},\ \bibinfo {pages} {221} (\bibinfo {year} {2013})}\BibitemShut
  {NoStop}%
\bibitem [{\citenamefont {H\"ubner}\ and\ \citenamefont {Zhang}(1998)}]{HZ98}%
  \BibitemOpen
  \bibfield  {author} {\bibinfo {author} {\bibfnamefont {W.}~\bibnamefont
  {H\"ubner}}\ and\ \bibinfo {author} {\bibfnamefont {G.~P.}\ \bibnamefont
  {Zhang}},\ }\href {\doibase 10.1103/PhysRevB.58.R5920} {\bibfield  {journal}
  {\bibinfo  {journal} {Phys. Rev. B}\ }\textbf {\bibinfo {volume} {58}},\
  \bibinfo {pages} {R5920} (\bibinfo {year} {1998})}\BibitemShut {NoStop}%
\bibitem [{\citenamefont {Bhattacharjee}\ \emph {et~al.}(2012)\citenamefont
  {Bhattacharjee}, \citenamefont {Nordstr\"om},\ and\ \citenamefont
  {Fransson}}]{BNF12}%
  \BibitemOpen
  \bibfield  {author} {\bibinfo {author} {\bibfnamefont {S.}~\bibnamefont
  {Bhattacharjee}}, \bibinfo {author} {\bibfnamefont {L.}~\bibnamefont
  {Nordstr\"om}}, \ and\ \bibinfo {author} {\bibfnamefont {J.}~\bibnamefont
  {Fransson}},\ }\href {\doibase 10.1103/PhysRevLett.108.057204} {\bibfield
  {journal} {\bibinfo  {journal} {Phys. Rev. Lett.}\ }\textbf {\bibinfo
  {volume} {108}},\ \bibinfo {pages} {057204} (\bibinfo {year}
  {2012})}\BibitemShut {NoStop}%
\bibitem [{\citenamefont {Mankovsky}\ \emph {et~al.}(2022)\citenamefont
  {Mankovsky}, \citenamefont {Polesya}, \citenamefont {Lange}, \citenamefont
  {Wei\ss{}enhofer}, \citenamefont {Nowak},\ and\ \citenamefont
  {Ebert}}]{MPL+22}%
  \BibitemOpen
  \bibfield  {author} {\bibinfo {author} {\bibfnamefont {S.}~\bibnamefont
  {Mankovsky}}, \bibinfo {author} {\bibfnamefont {S.}~\bibnamefont {Polesya}},
  \bibinfo {author} {\bibfnamefont {H.}~\bibnamefont {Lange}}, \bibinfo
  {author} {\bibfnamefont {M.}~\bibnamefont {Wei\ss{}enhofer}}, \bibinfo
  {author} {\bibfnamefont {U.}~\bibnamefont {Nowak}}, \ and\ \bibinfo {author}
  {\bibfnamefont {H.}~\bibnamefont {Ebert}},\ }\href {\doibase
  10.1103/PhysRevLett.129.067202} {\bibfield  {journal} {\bibinfo  {journal}
  {Phys. Rev. Lett.}\ }\textbf {\bibinfo {volume} {129}},\ \bibinfo {pages}
  {067202} (\bibinfo {year} {2022})}\BibitemShut {NoStop}%
\bibitem [{\citenamefont {Mankovsky}\ \emph {et~al.}(2023)\citenamefont
  {Mankovsky}, \citenamefont {Lange}, \citenamefont {Polesya},\ and\
  \citenamefont {Ebert}}]{MLPE23}%
  \BibitemOpen
  \bibfield  {author} {\bibinfo {author} {\bibfnamefont {S.}~\bibnamefont
  {Mankovsky}}, \bibinfo {author} {\bibfnamefont {H.}~\bibnamefont {Lange}},
  \bibinfo {author} {\bibfnamefont {S.}~\bibnamefont {Polesya}}, \ and\
  \bibinfo {author} {\bibfnamefont {H.}~\bibnamefont {Ebert}},\ }\href
  {\doibase 10.1103/PhysRevB.107.144428} {\bibfield  {journal} {\bibinfo
  {journal} {Phys. Rev. B}\ }\textbf {\bibinfo {volume} {107}},\ \bibinfo
  {pages} {144428} (\bibinfo {year} {2023})}\BibitemShut {NoStop}%
\bibitem [{ELK()}]{ELK}%
  \BibitemOpen
  \href@noop {} {\enquote {\bibinfo {title} {{The Elk Code}},}\ }\bibinfo
  {howpublished} {\url{http://elk.sourceforge.net/}}\BibitemShut {NoStop}%
\bibitem [{\citenamefont {Liechtenstein}\ \emph {et~al.}(1987)\citenamefont
  {Liechtenstein}, \citenamefont {Katsnelson}, \citenamefont {Antropov},\ and\
  \citenamefont {Gubanov}}]{LKAG87}%
  \BibitemOpen
  \bibfield  {author} {\bibinfo {author} {\bibfnamefont {A.~I.}\ \bibnamefont
  {Liechtenstein}}, \bibinfo {author} {\bibfnamefont {M.~I.}\ \bibnamefont
  {Katsnelson}}, \bibinfo {author} {\bibfnamefont {V.~P.}\ \bibnamefont
  {Antropov}}, \ and\ \bibinfo {author} {\bibfnamefont {V.~A.}\ \bibnamefont
  {Gubanov}},\ }\href {\doibase 10.1016/0304-8853(87)90721-9} {\bibfield
  {journal} {\bibinfo  {journal} {J. Magn. Magn. Materials}\ }\textbf {\bibinfo
  {volume} {67}},\ \bibinfo {pages} {65} (\bibinfo {year} {1987})}\BibitemShut
  {NoStop}%
\bibitem [{\citenamefont {Udvardi}\ \emph {et~al.}(2003)\citenamefont
  {Udvardi}, \citenamefont {Szunyogh}, \citenamefont {Palot\'as},\ and\
  \citenamefont {Weinberger}}]{USPW03}%
  \BibitemOpen
  \bibfield  {author} {\bibinfo {author} {\bibfnamefont {L.}~\bibnamefont
  {Udvardi}}, \bibinfo {author} {\bibfnamefont {L.}~\bibnamefont {Szunyogh}},
  \bibinfo {author} {\bibfnamefont {K.}~\bibnamefont {Palot\'as}}, \ and\
  \bibinfo {author} {\bibfnamefont {P.}~\bibnamefont {Weinberger}},\ }\href
  {\doibase 10.1103/PhysRevB.68.104436} {\bibfield  {journal} {\bibinfo
  {journal} {Phys. Rev. B}\ }\textbf {\bibinfo {volume} {68}},\ \bibinfo
  {pages} {104436} (\bibinfo {year} {2003})}\BibitemShut {NoStop}%
\bibitem [{\citenamefont {Ebert}\ and\ \citenamefont
  {Mankovsky}(2009)}]{EM09a}%
  \BibitemOpen
  \bibfield  {author} {\bibinfo {author} {\bibfnamefont {H.}~\bibnamefont
  {Ebert}}\ and\ \bibinfo {author} {\bibfnamefont {S.}~\bibnamefont
  {Mankovsky}},\ }\href {\doibase 10.1103/PhysRevB.79.045209} {\bibfield
  {journal} {\bibinfo  {journal} {Phys. Rev. B}\ }\textbf {\bibinfo {volume}
  {79}},\ \bibinfo {pages} {045209} (\bibinfo {year} {2009})}\BibitemShut
  {NoStop}%
\bibitem [{\citenamefont {Mankovsky}\ \emph {et~al.}(2020)\citenamefont
  {Mankovsky}, \citenamefont {Polesya},\ and\ \citenamefont {Ebert}}]{MPE20}%
  \BibitemOpen
  \bibfield  {author} {\bibinfo {author} {\bibfnamefont {S.}~\bibnamefont
  {Mankovsky}}, \bibinfo {author} {\bibfnamefont {S.}~\bibnamefont {Polesya}},
  \ and\ \bibinfo {author} {\bibfnamefont {H.}~\bibnamefont {Ebert}},\ }\href
  {\doibase 10.1103/PhysRevB.101.174401} {\bibfield  {journal} {\bibinfo
  {journal} {Phys. Rev. B}\ }\textbf {\bibinfo {volume} {101}},\ \bibinfo
  {pages} {174401} (\bibinfo {year} {2020})}\BibitemShut {NoStop}%
\bibitem [{\citenamefont {Staunton}\ \emph {et~al.}(2006)\citenamefont
  {Staunton}, \citenamefont {Szunyogh}, \citenamefont {Buruzs}, \citenamefont
  {Gyorffy}, \citenamefont {Ostanin},\ and\ \citenamefont {Udvardi}}]{SSB+06}%
  \BibitemOpen
  \bibfield  {author} {\bibinfo {author} {\bibfnamefont {J.~B.}\ \bibnamefont
  {Staunton}}, \bibinfo {author} {\bibfnamefont {L.}~\bibnamefont {Szunyogh}},
  \bibinfo {author} {\bibfnamefont {A.}~\bibnamefont {Buruzs}}, \bibinfo
  {author} {\bibfnamefont {B.~L.}\ \bibnamefont {Gyorffy}}, \bibinfo {author}
  {\bibfnamefont {S.}~\bibnamefont {Ostanin}}, \ and\ \bibinfo {author}
  {\bibfnamefont {L.}~\bibnamefont {Udvardi}},\ }\href {\doibase
  10.1103/PhysRevB.74.144411} {\bibfield  {journal} {\bibinfo  {journal} {Phys.
  Rev. B}\ }\textbf {\bibinfo {volume} {74}},\ \bibinfo {pages} {144411}
  (\bibinfo {year} {2006})}\BibitemShut {NoStop}%
\bibitem [{\citenamefont {Ebert}\ \emph {et~al.}(2015)\citenamefont {Ebert},
  \citenamefont {Mankovsky}, \citenamefont {Chadova}, \citenamefont {Polesya},
  \citenamefont {Min\'{a}r},\ and\ \citenamefont {K\"odderitzsch}}]{EMC+15}%
  \BibitemOpen
  \bibfield  {author} {\bibinfo {author} {\bibfnamefont {H.}~\bibnamefont
  {Ebert}}, \bibinfo {author} {\bibfnamefont {S.}~\bibnamefont {Mankovsky}},
  \bibinfo {author} {\bibfnamefont {K.}~\bibnamefont {Chadova}}, \bibinfo
  {author} {\bibfnamefont {S.}~\bibnamefont {Polesya}}, \bibinfo {author}
  {\bibfnamefont {J.}~\bibnamefont {Min\'{a}r}}, \ and\ \bibinfo {author}
  {\bibfnamefont {D.}~\bibnamefont {K\"odderitzsch}},\ }\href {\doibase
  http://dx.doi.org/10.1103/PhysRevB.91.165132} {\bibfield  {journal} {\bibinfo
   {journal} {Phys. Rev. B}\ }\textbf {\bibinfo {volume} {91}},\ \bibinfo
  {pages} {165132} (\bibinfo {year} {2015})}\BibitemShut {NoStop}%
\bibitem [{\citenamefont {Ebert}\ \emph {et~al.}(2016)\citenamefont {Ebert},
  \citenamefont {Braun}, \citenamefont {K\"odderitzsch},\ and\ \citenamefont
  {Mankovsky}}]{EBKM16}%
  \BibitemOpen
  \bibfield  {author} {\bibinfo {author} {\bibfnamefont {H.}~\bibnamefont
  {Ebert}}, \bibinfo {author} {\bibfnamefont {J.}~\bibnamefont {Braun}},
  \bibinfo {author} {\bibfnamefont {D.}~\bibnamefont {K\"odderitzsch}}, \ and\
  \bibinfo {author} {\bibfnamefont {S.}~\bibnamefont {Mankovsky}},\ }\href
  {\doibase 10.1103/PhysRevB.93.075145} {\bibfield  {journal} {\bibinfo
  {journal} {Phys. Rev. B}\ }\textbf {\bibinfo {volume} {93}},\ \bibinfo
  {pages} {075145} (\bibinfo {year} {2016})}\BibitemShut {NoStop}%
\end{thebibliography}
%\bibliographystyle{apsrev4-1}


\end{document}